\documentclass[nofootinbib, superscriptaddress, twocolumn]{revtex4-2}
\usepackage{graphicx} 
\usepackage{graphicx} 
\usepackage{tikz}
\usetikzlibrary{arrows}
\usepackage{multirow}
\usepackage[colorlinks=true,urlcolor=blue,linkcolor=blue,citecolor=blue]{hyperref}
\usepackage[normalem]{ulem}
\usepackage{amsmath}
\usepackage[capitalize]{cleveref}
\usepackage{amsfonts,amssymb}

\usepackage{comment}
\usepackage{flushend}

\usepackage[T1]{fontenc} 
\graphicspath{{fig/}}
\usepackage{mathtools}
\usepackage{xcolor}
\usepackage{graphicx}
\usepackage{subfigure}
\usepackage{dcolumn}
\usepackage{bm}

\def\dd{\mathrm{d}}

\definecolor{myblue}{RGB}{0, 100, 200}
\definecolor{myred}{RGB}{214, 39, 40}
\definecolor{mygreen}{RGB}{44, 160, 44}
\definecolor{mybrown}{RGB}{123, 64, 26}
\definecolor{mydarkblue}{RGB}{44, 77, 118}

\begin{document}

\title{Extracting Properties of Dark Dense Environments around Black Holes from Gravitational Waves}

\author{Qianhang Ding}
\email{dingqh@ibs.re.kr}
\affiliation{Cosmology, Gravity and Astroparticle Physics Group, Center for Theoretical Physics of the Universe,
Institute for Basic Science (IBS), Daejeon, 34126, Korea}

\author{Minxi He}
\email{heminxi@ibs.re.kr}
\affiliation{Particle Theory and Cosmology Group, Center for Theoretical Physics of the Universe,
Institute for Basic Science (IBS), Daejeon, 34126, Korea}

\author{Hui-Yu Zhu}
\email{hzhuav@ibs.re.kr}
\affiliation{Cosmology, Gravity and Astroparticle Physics Group, Center for Theoretical Physics of the Universe,
Institute for Basic Science (IBS), Daejeon, 34126, Korea}

\begin{abstract}
    Dark matter (DM) can form dense condensates around black holes (BHs), such as superradiant clouds and ultracompact mini halos, which can significantly affect the orbital evolution of their companion objects through dynamical friction (DF). 
    In this work, we define a novel quantity to quantify such effects in the emitted gravitational waves (GWs) in terms of GW amplitude, frequency, and their time derivatives. 
    The information about the density profile can be extracted from this quantity, which characterizes the type of condensate and, therefore, the corresponding DM property. 
    This quantity allows us to probe the dark dense environment by multi-wavelength GW observation with existing ground-based and future space-based GW detectors, potentially revealing the properties of the dark sector and shedding light on the primordial origin of the stellar mass BHs. 
    A null detection can place strong constraints on the relevant DM parameters. 
\end{abstract}

\maketitle

\section{Introduction}

Since the detection of the first gravitational wave (GW) event, GW150914, from a binary black hole (BBH) merger \cite{LIGOScientific:2016aoc}, observations of BBH mergers have become a powerful tool for studying black hole (BH) physics and probing fundamental physics in the strong gravity regime. 
Such detections enable systematic tests of general relativity \cite{LIGOScientific:2020tif}, the no-hair theorem \cite{Isi:2019aib}, and Hawking’s area law \cite{Isi:2020tac, Tang:2025jyj, KAGRA:2025oiz}, etc. 
The BH parameters and luminosity distance extracted from GW waveforms also make BBH a standard siren for studying cosmic evolution \cite{Schutz:1986gp, Holz:2005df, LIGOScientific:2017vwq, Ding:2023smy, Chen:2024gdn}. 
Moreover, the mass distribution of these events provides potential evidence for the existence of primordial black holes (PBHs) \cite{Clesse:2020ghq, DeLuca:2020sae, Yuan:2025avq}. 
These studies motivate further exploration of new physics accessible through GW observations of BBHs.

The LIGO–Virgo–KAGRA (LVK) collaboration can observe the GW of inspiral phase, merger phase and ringdown phase of mergers of BBHs and other compact objects such as neutron stars. 
The GW in  ringdown phase carries rich information about the fundamental aspects of gravity and BHs (see e.g.~\cite{Kokkotas:1999bd,Berti:2009kk,Konoplya:2011qq}). 
On the other hand, the GW in inspiral phase offers a channel to test the ambient environments of BHs~\cite{Chen:2020lpq, Takatsy:2025bfk}, where dark matter (DM) and beyond Standard Model (BSM) particles could play a crucial role in affecting the GW waveforms.
Due to the strong gravitational potential of BHs, DM can be gravitationally captured and form a dense structure around the BHs \cite{Gondolo:1999ef, Eda:2013gg, Bar:2019pnz, Mack:2006gz}, which makes the vicinity of BHs an ideal laboratory to study DM. Various methods have been proposed to study DM in such scenarios \cite{Akil:2023kym, Kadota:2023wlm, Aurrekoetxea:2023jwk, Ding:2024mro, Tomaselli:2025zdo, Spieksma:2025exm, Roy:2025qaa}. The existence of dense DM in the BH surrounding significantly affects the GW waveform during the inspiral phase \cite{Eda:2014kra, Ghodla:2024fit}. Therefore, precise observations of GW waveforms provide a unique probe of DM physics in the strong-gravity regime.

The dark dense environments can be classified according to their density profiles, which typically correspond to different formation mechanisms. For instance, around a spinning black hole, the presence of a DM seed can lead to the formation of a bosonic cloud, as the black hole loses its angular momentum. 
The cloud–BH system is  
regarded as a gravitational atom (GA), and the cloud growth phenomenon is known as superradiant instability \cite{Arvanitaki:2009fg,Arvanitaki:2010sy,Brito:2015oca,Baumann:2019eav}. 
Once 
GA is formed, it can influence the orbital motion of its binary companion through several mechanisms. 
For instance, state transitions and couplings between the eigenstates of the GA can induce backreaction on the companion \cite{Baumann:2019ztm,Ding:2020bnl,Takahashi:2021yhy,Tong:2021whq,Tong:2022bbl,Baumann:2021fkf,Takahashi:2023flk,Tomaselli:2023ysb,Fan:2023jjj,Zhu:2024bqs}. 
In addition, when the companion is located inside the cloud, dynamical friction (DF) leads to extra power loss 
of kinetic energy, significantly speeding up the orbital shrinkage~\cite{Baumann:2021fkf,Tomaselli:2023ysb,Dosopoulou:2023umg,Ding:2025nxe}. 
Hence, the information of this boson cloud, such as boson mass, can be potentially inferred from GW signals.

Another commonly discussed dense environment formed around a BH by gravitational attraction is compact mini halos~\cite{Gondolo:1999ef, Gnedin:2003rj, Bar:2019pnz, Eda:2013gg, Mack:2006gz}.
Their formation requires either a supermassive black hole (SMBH) or long-term accumulation of DM particles around PBHs \cite{Mack:2006gz, Ricotti:2007au, Berezinsky:2013fxa}. 
Therefore, the detection of a DM halo surrounding a stellar-mass BH could indicate its primordial origin \cite{Oguri:2022fir, GilChoi:2023ahp}. 
The density profile of DM halos around BHs depends on the nature of DM. 
Ultralight bosons would form a soliton-like DM halo around a Schwarzschild BH \cite{Bar:2019pnz, Brax:2019npi, Hui:2019aqm}, while the particle-like cold dark matter (CDM) typically forms a spike-like distribution around BHs. 
The density profile of DM halos can affect the orbital evolution of the companion by DF and leave imprints on the emitted GWs, which we aim to obtain from the waveform of GWs emitted during the inspiral phase of binary mergers.

In this work, we mainly focus on BBH systems, and introduce a new quantity, denoted as $D$, which is the combination of three characteristics in a GW signal: the amplitude $h$, the frequency $f$, and its time derivative ${\rm d}f/{\rm d}t$.
$D$ is proportional to the additional power loss caused by DF, and its magnitude depends on the density of the dark environment experienced by the binary companion at a given GW frequency. 
Therefore, analyzing the evolution of $D$ of a BBH GW signal in the frequency domain allows one to reconstruct the density profile of the dark environment around those BHs. 
We show that fitting the theoretical $D-f$ relation for future GW observation signals can constrain the mass of ultralight boson in the superradiant cloud, the soliton structure around BHs, and the power-law index of DM halo density profiles. 
In the following discussions, we will consider circular orbits for two main reasons. The first is to simplify the analysis and emphasize the main point of our method. The second and more practical one is that current observation is based on templates which are calculated for circular orbits. This is justified, because the eccentricity at the final stage of mergers has been significantly reduced by GW emission. Also, DF becomes relevant typically at small separation where circularization is expected.

This paper is organized as follows. 
In Sec.~\ref{sec:dark_dense_environment}, we introduce two types of important dark dense environments around BHs, GA boson clouds and DM halos. 
In Sec.~\ref{sec:GW_probe}, we construct a new quantity $D$ from the properties of GW waveforms and demonstrate the relation between the $D-f$ diagram and the density profile of the dark dense environment. 
In Sec.~\ref{sec:DM_detectability}, we show the frequency range of GWs within which $D$ can be efficiently measured, and how $D$ can be used to constrain DM properties in these regimes.
In Sec.~\ref{sec:conclusion}, we summarize how this unique quantity $D$ probes the nature of dark dense environment from GW waveforms.

\section{Dark dense environment}\label{sec:dark_dense_environment}

DM occupies about $80\%$ of the total matter density in the Universe \cite{Planck:2018vyg}.
Such a large fraction makes DM playing a significant role in forming the environments around compact objects, such as BHs. 
The presence of a dense dark environment around BHs can affect the orbital evolution of BBH systems and, therefore, modify the waveform of emitted GWs.

We mainly focus on two typical macro-structures formed around BHs in this paper: GAs and DM halos. 
In the case of a GA, the most widely discussed formation channel is superradiant instability, which suggests that a boson cloud can form around the BH by extracting the BH spin, and the mass of the cloud can reach at most $\sim \mathcal O(10\%)$ of the BH mass \cite{Tsukada:2018mbp}. 
On the other hand, DM halos form by the accretion of CDM over a long timescale.
The density profile of DM halos depends on DM models. 
For example, ultralight bosons can form a soliton-like DM halo, while 
particle CDM tends to have a spiky profile.
Such so-called “dressed” BHs could have a primordial origin \cite{Oguri:2022fir}.

The dark, dense environment around BBHs can provide an additional dissipation channel, DF \cite{1943ApJ....97..255C}, to the orbital energy besides GW emission. The power of DF can be described by
\begin{align}\label{eq:DF_power}
    P_{\rm DF} = - 4 \pi \frac{G^2 M_\ast^2}{v} \rho_{D} C_\Lambda ~ ,
\end{align}
and it can dominate over GW emission power within a certain frequency range. Here, $M_\ast$ is the mass of the binary companion, $v$ is the relative velocity between the companion and the dark environment,
$\rho_D$ is the density of dark environment that the companion experiences, and $C_\Lambda$ is the Coulomb logarithm. 

The form of $C_\Lambda$ depends on the nature of DM. 
For ultralight boson with mass $ \mu $, $C_\Lambda$ depends on the parameter $\xi \equiv M_\ast \mu/v $. In the regime $\xi \ll 1$, $C_\Lambda$ can be approximated as \cite{Hui:2016ltb}
\begin{align}
    C_{\Lambda} \simeq \mathrm{Cin}(2kr_\Lambda) + \frac{\sin 2kr_\Lambda}{2kr_\Lambda} - 1  ~,
    \label{Eq.ULBCLambda}
\end{align}
where $k = \mu v$, and $r_\Lambda$ represents the smaller quantity between the size of the binary orbit and the typical scale of the dense environment. 
Function $\text{Cin}$ is defined as $\mathrm{Cin}(z) = \int_0^z (1 - \cos y)/y \, \dd y$.
When $ kr_\Lambda \to 0 $, $ C_{\Lambda} \to (kr)^2/3 $.
For particle-like CDM, the Coulomb logarithm can be expressed in a classical form in Chandrasekhar's formula \cite{1943ApJ....97..255C} as
\begin{align}
    C_\Lambda \simeq \ln \Lambda~,
\end{align}
where $\Lambda \equiv v^2 r/G M_\ast $ with mass ratio $q \equiv M_* / M $ between the companion and the dressed BH. 
For circular orbit, $ \Lambda= (1+q)/q $, so $C_\Lambda$ is solely fixed by $ q $ in this case.

It is worth noting that throughout this work, we assume the orbital shrinking proceeds adiabatically. However, strictly speaking, the relative velocity in Eq.~\eqref{eq:DF_power} contains both radial and 
angular components during inspiraling, $\vec{v} = v_r \hat{r} + v_\phi \hat{\phi}$, where $\hat{r}$ and $\hat{\phi}$ are the unit vectors for these two directions, respectively.  
The contribution from $v_r$ can be safely neglected only when $|v_\phi| \gg |v_r|$. 
Under the circular orbit approximation, this condition can be estimated as $|v_r / v_\phi| \sim -\,\dd (f^{-1})/\dd t \ll 1$.

This condition can be further expressed in terms of the binary separation $r$ and the ratio $|P_{\rm DF}/P_{\rm GW}|$, providing an analytical criterion for self-consistency. In this work, we verify it numerically for each case and identify the valid regimes in the examples discussed below. The detailed calculation of deriving the consistency check criterion for the adiabatic orbital shrinking assumption is presented in Appendix~\ref{Sec.selfconsistency}.

Next, we will introduce the dark dense environment we mentioned above in detail, and calculate their density profiles.

\subsection{Superradiant Instability and GA}\label{Sec.SuperradiantBackground}
In this section, we will provide a brief overview of the background information regarding superradiant instability as outlined in reference \cite{Baumann:2019ztm}.

Superradiant instability refers to the phenomenon where a rotating BH with mass $M$ can create a bosonic cloud through the loss of angular momentum, leading to the formation of a structure known as a GA. 
This condition requires the boson's Compton wavelength to be larger than the Schwarzschild radius of the BH. 
The amplification behavior can be understood by solving the Klein-Gordon (KG) equation within a Kerr background 
\begin{equation}\label{eq:KG_eq}
    \Big(\Box_{\rm Kerr}-\mu^{2}\Big) \Phi=0 ~ , 
\end{equation}
where $ \Phi $ is the bosonic field. 
After taking into account the ingoing boundary condition at the horizon, along with the decaying boundary condition at infinity, we can derive a series of eigenmodes characterized by three quantum numbers denoted as $|nlm\rangle$. 
The corresponding eigenvalues can be numerically solved using Leaver's method \cite{Teukolsky:1973ha,Leaver:1985ax}, and their analytical form can be obtained in the non-relativistic limit. It is noteworthy that the eigenfrequency in the solution contains an 
imaginary part $\omega_{nlm}=E_{nlm}+i\Gamma_{nlm}$, which results in the growth of the cloud. 
The growth can be observed from the amplitude of the mode function 
$\psi_{nlm}\propto e^{i\omega_{nlm}t}\propto e^{\Gamma_{nlm} t}$. 
Here, the rate of superradiance $\Gamma_{nlm}$ has an analytical form \cite{Detweiler:1980uk} 
\begin{equation}
    \Gamma_{n\ell m} = 2\,\tilde r_{+}\,C_{n\ell}\,g_{\ell m}\bigl(m\Omega_{\rm BH}-\omega_{n\ell m}\bigr)\,\alpha^{4\ell+5} ~ ,
    \label{Eq.SuperRadRate}
\end{equation}
where $\tilde r_+$ denotes the outer horizon and $\Omega_{\rm BH}$ is the angular velocity of the Kerr BH. 
$\alpha \equiv GM\mu/(c\hbar)$ is the gravitational fine-structure constant, representing the ratio of the Schwarzschild radius of the BH to the Compton wavelength of the bosonic field, with $G$, $c$, and $\hbar$ representing the gravitational constant, speed of light, and reduced Planck constant, respectively. 
Note that the analytical expression is valid only for $\alpha < 0.3$ \cite{Brito2015b, Cannizzaro:2023jle}. The explicit forms of $C_{nl}$ and $g_{lm}$ can be found in \cite{Detweiler:1980uk, Baumann:2018vus, Baumann:2019eav}.

From Eq.~\eqref{Eq.SuperRadRate}, we see that exponential growth occurs only when 
$ {\rm Re}\,(m\Omega_{\rm BH} - \omega_{nlm}) >0 $, which requires a highly-spinning black hole. 
Otherwise, $\Gamma_{nlm}$ becomes negative, and the cloud amplitude decays exponentially with time, indicating that the cloud will be absorbed by the BH. 
Moreover, if the initial spin is sufficiently high, the mode $|211\rangle$ will dominate over other states, since it has the lowest orbital quantum number $l$, and the superradiant rate is strongly suppressed by $\alpha^{4l+5}$ for $\alpha \ll 1$.

Together with the eigenfrequency, we can derive the density of the cloud \cite{Cao:2023fyv,Li:2025qyu,Ding:2025nxe}
\begin{align}\label{eq:density_profile} 
    \rho_{nlm}(x, \theta) =& ~1.01 \times 10^{30} \, M_\odot \,\mathrm{pc^{-3}}   \notag\\ &\times  A_{nlm} (x, \theta) \frac{\beta}{0.1} \left(\frac{\alpha}{0.1}\right)^6 \left(\frac{30 \,M_\odot}{M_{B}}\right)^2~,   
\end{align}
where $A_{nlm}(x, \theta)$ is the spatial density distribution, with $x = r/r_0$, with $ r_0 \equiv 1/\mu\alpha $ denotes the Bohr radius of GA. Here $\beta$ is the mass ratio between the boson cloud and BH.
For the most dominant mode $|211\rangle$, we have 
\begin{equation}
    A_{211}(x,\theta)= \frac{1}{64\pi}   x^{2}e^{-x} \sin^{2}\theta ~ .
    \label{Eq.A211}
\end{equation}

\subsection{Dark Matter Halo}

The BH gravitational attraction would accumulate DM and form a dense halo around it \cite{Gondolo:1999ef,  Bar:2019pnz, Mack:2006gz, Eda:2013gg}. 
The formation of such a DM halo requires either a strong gravitational potential, as provided by an SMBH \cite{Gondolo:1999ef, Gnedin:2003rj}, or a sufficiently long accumulation timescale, which is likely to be achieved in the case of a PBH \cite{Mack:2006gz, Ricotti:2007au}. 
The profile of the halo depends on DM model. Wave-like DM forms soliton-like structure, while particle-like CDM, 
such as weakly interacting massive particles (WIMP), forms a spike-like structure.

For ultralight bosons around a Schwarzschild BH, a soliton-like DM halo would form \cite{Bar:2019pnz, Brax:2019npi} with a density profile follows \cite{Aghaie:2023lan}
\begin{align}\label{eq:soliton_density_profile}\nonumber
    \rho_{\rm sol}(x) =& ~4.05 \times 10^{26} \, M_\odot \, \text{pc}^{-3} \\
    &\times \varepsilon  e^{-2x} \left(\frac{M}{30 \, M_\odot}\right)^4 \left(\frac{\mu}{10^{-13} \, \text{eV}}\right)^6~,
\end{align}
where $\varepsilon$ is the mass ratio between the solitonic core and BH. Note that although we are discussing the macro-structure consisting of the ultralight boson, its density profile is different from the GA cloud density profile given in Eq.~\eqref{eq:density_profile}. Because in the context of superradiance, we focus on spinning Kerr BHs, while for the soliton case we consider a non-rotating BH, around which the ultralight bosons can only condense without extracting additional energy from the BH \cite{Hui:2019aqm}. 

In the case of particle-like CDM, it can undergo gravitational collapse to form a DM halo with an inner cuspy profile even in the absence of a BH, which is typically modeled by the Navarro–Frenk–White (NFW) profile \cite{Navarro:1995iw, Navarro:1996gj}. If an SMBH appears at the center of the galactic DM halo, its strong gravitational field causes the surrounding DM growing adiabatically. Ref.~\cite{Gondolo:1999ef} proposed that this process leads to the formation of a spike-like dense DM environment around SMBHs, with a density profile given by
\begin{align}
    \rho_{\rm sp} (r) = \rho_R \left(1 - \frac{4 r_s}{r}\right)^3 \left(\frac{r_{\rm sp}}{r}\right)^\gamma~,
\end{align}
where $\rho_R$ is the density at the boundary of the spike, $r_{\rm sp}$ is the radius of the DM spike, and $r_s$ is the Schwarzschild radius of the SMBH. The existence of such a spike-like DM halo would cause various potential observational signals, such as DM annihilation \cite{Gondolo:1999ef} and PBH mergers \cite{Nishikawa:2017chy, Ding:2024mro}.

On the other hand, for stellar-mass BHs, their gravitational potential is not strong enough to significantly affect a DM halo with the NFW profile. 
However, a spike-like DM halo can still form around them through long-term accumulation.  
In this scenario, an astrophysical BH is unlikely to host such a DM halo due to its young age. 
In contrast, a PBH formed in the early Universe can accrete DM over the entire cosmic lifetime, leading to the formation of a dense halo around it \cite{Mack:2006gz, Ricotti:2007au, Berezinsky:2013fxa}. 
The density profile of DM halo surrounding a PBH can be expressed as \cite{Berezinsky:2013fxa, Boudaud:2021irr}
\begin{align}
    \rho_h(r) = \rho_0 \left(\frac{r_h}{r}\right)^{9/4} ~ ,
\end{align}
where $\rho_0=0.26 \, M_\odot \, \text{pc}^{-3} ((1+z_c)/31)^3$ is the density at the boundary of the DM halo. 
The typical halo size is characterized by $r_h = 0.61 \, \text{pc} (31/(1+z_c)) (M_h/M_\odot)^{1/3}$ with a mass $M_h = 97\times(31/(1+z_c)) M$. 
$z_c$ is the cutoff redshift for DM accretion, which is around $z\sim 30$. 
In general, the density profile of a CDM halo surrounding a BH follows $\rho \propto r^{\gamma}$, characterized by index $ \gamma $.

\section{A dark dense environment probe in GWs}\label{sec:GW_probe}

GW emission dissipates the total energy of BH binary system, leading to
the orbital shrinkage and the increase of GW frequency. 
Meanwhile, the presence of a dense dark environment around BHs introduces additional energy dissipation through DF, which further affects the evolution of orbital motion and, therefore, GW frequency. 
Consequently, the non-standard evolution of GW frequency can serve as a smoking gun for the existence of dark environment and an effective probe of its properties.

Let us consider a system in which a binary companion is inspiraling around a BH embedded in a dense dark environment. 
The frequency evolution caused by GW emission and DF dissipation can be expressed as
\begin{align}\label{eq:frequency_evo_source}
    \frac{\dd f_s}{\dd t_s} = - \frac{3}{(\pi G)^{2/3}} \frac{f_s^{1/3}}{\mathcal{M}_c^{5/3}} (P_{\rm GW} + P_{\rm DF})~.
\end{align}
Here $f_s$ and $\dd t_s$ are the intrinsic frequency and time interval in the source frame, which differ from those measured on Earth, $f$ and $\dd t$, by  
\begin{equation}
        f_s = (1+z)f ~ ,\quad \dd t_s  = (1+z)^{-1}\dd t ~ . 
        \label{Eq.SourceObserveTrans}
\end{equation}
Here $z$ is the cosmic redshift. 
$\mathcal{M}_c$ is the chirp mass of the BH binary, defined in terms of the BH mass $M$ and the companion mass $M_*$ as $\mathcal{M}_c \equiv (M M_*)^{3/5}/(M + M_*)^{1/5}$. $P_{\rm GW}$ and $P_{\rm DF}$ denote the powers of GW emission and DF, respectively. For a circular orbit, $P_{\rm GW}$ reads \cite{Peters:1964zz}
\begin{equation}\label{eq:GW_power}
     P_{\rm GW}=-\frac{32}{5}\frac{G^{7/3}}{c^5} (\pi \mathcal{M}_c f_s)^{10/3} ~ .  
\end{equation}
In this work, we focus on the case of a circular orbit $e = 0$. 
This is a reasonable assumption, as GW emission tends to circularize the orbit, particularly during the final stage of the inspiral phase in a binary system \cite{Hinder:2007qu, Franciolini:2021xbq}.
This is also one of the main reasons why the waveform templates used by current GW experiments, such as those conducted by the LVK Collaboration, are computed under the assumption $e = 0$, since their sensitive frequency bands correspond to the merger phase of stellar-mass compact binaries.

\begin{figure}[htbp]
	\centering
	\includegraphics[width=8cm]{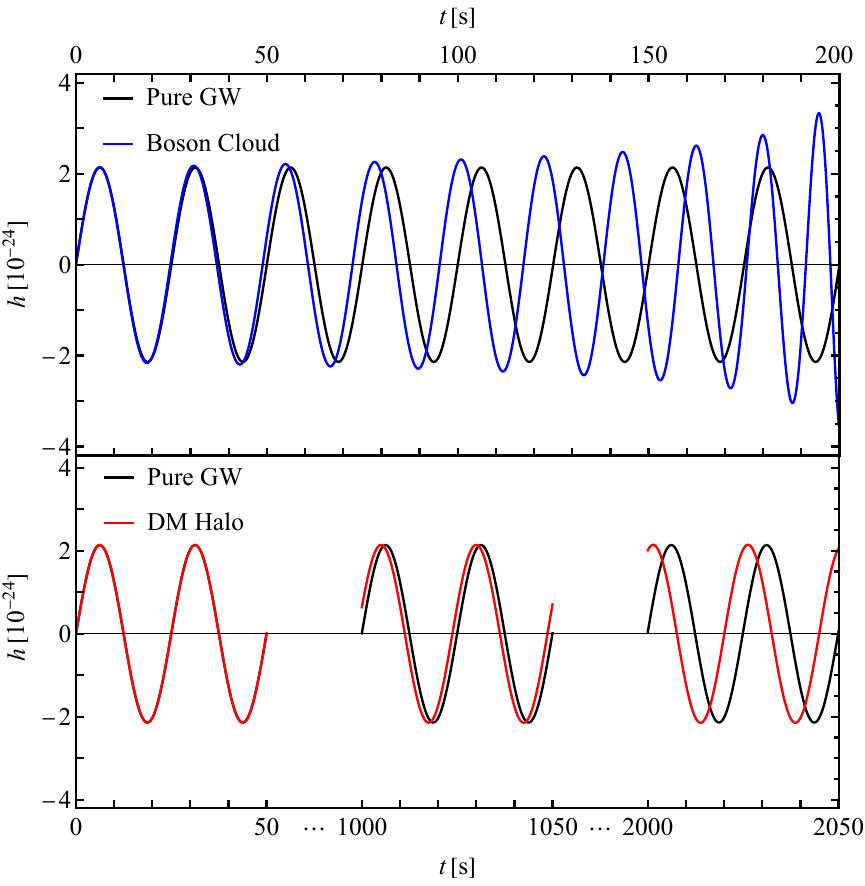}
	\caption{[Top] The GW waveform with and without the impacts from a superradiant boson cloud. We choose a $|211\rangle$ state boson cloud with $\alpha = 0.1$. [Bottom] The GW waveform with and without the impacts from a spike-like DM halo. We choose DM halo power index as $\gamma = -9/4$. The parameters of BH binary is set as $M = 30 \, M_\odot$, $q = 1$. The initial frequency is set as $f_{\rm ini} = 0.04 \, \text{Hz}$.}
	\label{fig:waveform}
\end{figure}

After transferring the intrinsic quantities into the ones observed on earth via the relation shown in Eq.~\eqref{Eq.SourceObserveTrans}, 
we apply Eq.~\eqref{eq:GW_power} to Eq.~\eqref{eq:frequency_evo_source} and obtain 
\begin{align}\label{eq:frequency_evo_obs}\nonumber
     \frac{\dd f}{\dd t} =& \frac{96}{5}\frac{[G\mathcal{M}_c(1+z)]^{5/3} \pi^{8/3} f^{11/3}}{c^5}\\
    &+\frac{3 f^{1/3}}{(\pi G)^{2/3} [\mathcal{M}_c(1+z)]^{5/3}}|P_{\rm DF}|  ~ ,
\end{align}
where the magnitude of the DF power, $|P_{\rm DF}|$, is calculated in the source frame of the BH binary. We then incorporate the DF power in various dense dark environments to illustrate its impact on the GW waveform in the observer’s rest frame, as shown in Fig.~\ref{fig:waveform}.
In this figure, we consider a BBH system with a component mass of $M = 30\,M_\odot$ and a mass ratio $q= 1$, and set the initial frequency of $f_{\rm ini} = 0.04~\text{Hz}$.
The upper panel shows the GW waveform dephasing in the presence of a GA cloud around the BH, in which we notice that the waveform changes rapidly. That is because when the DF induced by the GA cloud dominates over GW emission, the binary system is already close to the merger phase, where both the frequency $f$ and the GW power $P_{\rm GW}$, as well as the DF power $P_{\rm DF}$, are significantly larger. As indicated by Eq.~\eqref{eq:frequency_evo_source}, the time derivative $\dd f/\dd t$ tends to be large in this regime. In contrast, the lower panel shows the waveform dephasing in the presence of a spike-like DM halo. In this case, the dephasing proceeds more slowly but it becomes noticeable after a long accumulation time. This is because the DF induced by the spike-like DM halo dominates over GW emission only in the low-frequency regime, where the frequency evolution is correspondingly slower. 

In the following discussion, we aim at constructing an observable-dependent quantity that can reveal such a waveform difference, which reflects the contribution only from the DF power loss. 
Although we have 
the relation between $\dd f / \dd t$ and $P_{\rm DF}$ in Eq.~\eqref{eq:frequency_evo_obs}, this expression still involves parameters that must be determined through GW template fitting, such as the chirp mass $\mathcal{M}_c$. To obtain a cleaner form expressed purely in terms of observables, we introduce the relation between GW amplitude and its frequency, which is \cite{2007gwte.book.....M}
\begin{equation}\label{eq:amplitude}
     h=\frac{4\pi^{2/3}}{d_L(z)}\frac{[G\mathcal{M}_c(1+z)]^{5/3}}{c^4} f^{2/3} ~ . 
\end{equation}
Here, $h$ is the GW amplitude measured by detectors, $d_L(z) \equiv (1 + z) \int_0^z c/H(z')dz'$ is the luminosity distance, and $H(z)$ is the Hubble parameter
as a function of redshift.
To eliminate the $\mathcal{M}_c$ in the final quantity, we define 
\begin{align}\label{eq:gfunction}
    g (t) &\equiv \frac{1}{h f^3} \frac{\dd f}{\dd t} =\frac{24}{5}\frac{d_L(z)\pi^2}{c}+\frac{12 G}{c^4 d_L(z)}\frac{|P_{\rm DF}|}{ h^2 f^2} ~ ,
\end{align}
in which the first term $24 d_L(z)\pi^2/5 c$ is time-independent. From Eq.~\eqref{eq:gfunction}, we can see that if there is no dark dense environment, the second term vanishes, and $g(t)$ will remain a constant. 
Thus, we can remove the first term by taking time derivative 
\begin{align}
    \frac{\dd g}{\dd t} = \frac{12G}{c^4d_L(z)}\frac{|P_{\rm DF}|}{h^2 f^3} \frac{\dd f}{\dd t}\left(\frac{\dd \ln \rho}{\dd \ln f} + \frac{\dd \ln C_\Lambda}{\dd \ln f} - \frac{11}{3}\right)  .
\end{align}
After removing the GW power-related term in the equation, we move the observable $h$, $f$, and $\dd f/\dd t$ from the right-hand side to the left-hand side to construct a new quantity $D$ in GW signals as
\begin{align}\label{eq:observable_D}\nonumber
    D &\equiv -h^2 f^3 \frac{\dd g/\dd t}{\dd f/\dd t}=\frac{\dd h}{\dd t} + 3\frac{h}{f} \frac{\dd f}{\dd t} - \frac{h}{\dd f/\dd t}\frac{\dd^2 f}{\dd t^2}\\
    &=\frac{12G}{c^4d_L(z)}|P_{\rm DF}| \left( \frac{11}{3} -\frac{\dd \ln \rho}{\dd \ln f} - \frac{\dd \ln C_\Lambda}{\dd \ln f} \right) ~ ,
\end{align}
which is determined by the DF power loss $P_{\rm DF}$, related to the density of the dark environment at the companion’s position as the orbit shrinks. 

To show how to use $D$ to reveal the property of dark dense environment, we consider three concrete examples as we have discussed in Sec.~\ref{sec:dark_dense_environment}, 
i.e. the GA boson cloud, soliton and spike-like DM halo. 

\subsection{Probing GA Boson Cloud via GWs}

\begin{figure}[htbp]
	\centering
	\includegraphics[width=8cm]{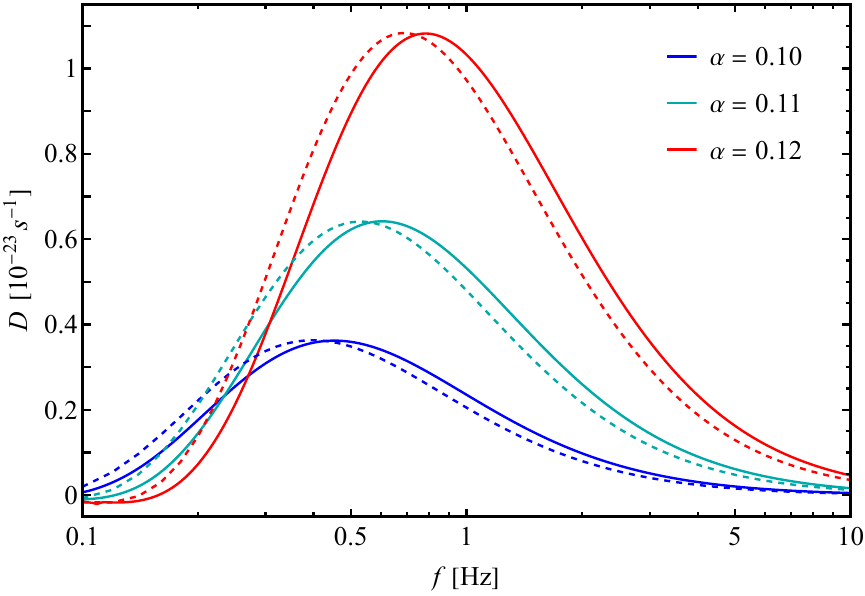}
	\caption{The $D-f$ diagram for a GA boson cloud with $\alpha = 0.1,\, 0.11,\, 0.12$, respectively. The solid curves represent the numerical results obtained using $C_\Lambda$ calculated from Eq.~\eqref{Eq.ULBCLambda}. The dashed curves show the best-fit analytical approximation of $D$, derived by assuming $C_\Lambda \propto f^{-2/3}$ in Eq.~\eqref{eq:D_boson_cloud}. The parameters of the BH binary are set to $M = 30\,M_\odot$ and $q = 1$, with the corresponding redshift fixed at $z = 0.5$.}
	\label{fig:D_BC}
\end{figure}

In this section, we discuss the DF induced by the GA bosonic cloud introduced in Sec.~\ref{Sec.SuperradiantBackground}. As an example, we focus on the dominant $|211\rangle$ mode and consider a binary orbit that lies in the equatorial plane, $\theta = \pi/2$. By taking  Eq.~\eqref{eq:density_profile} and \eqref{Eq.A211} into Eq.~\eqref{eq:DF_power} and \eqref{eq:observable_D}, we can then compute $D$ as 
\begin{align}\label{eq:D_boson_cloud}
    D  &= \frac{8G}{c^4d_L(z)} |P_{\rm DF}|\left(\frac{15}{2} - \left(\frac{f_0}{f}\right)^{\frac{2}{3}} - \frac{3}{2} \frac{\dd \ln C_\Lambda}{\dd \ln f} \right) \\\nonumber
    &\propto \left(\frac{f_0}{f}\right)^{\frac{5}{3}} e^{-\left(\frac{f_0}{f}\right)^{\frac{2}{3}}}\left(\frac{15}{2} - \left(\frac{f_0}{f}\right)^{\frac{2}{3}} - \frac{3}{2} \frac{\dd \ln C_\Lambda}{\dd \ln f} \right) C_\Lambda~.
\end{align}
Here, $f_0$ is the corresponding observed GW frequency when binary separation is the Bohr radius $r_0$. 
By fitting the $D-f$ relation in Eq.~\eqref{eq:D_boson_cloud} with the observed one, $f_0$ can be determined. 
Based on the Kepler's law, $ r_0 \sim M^{1/3} f_0^{-2/3} $ and therefore the corresponding boson mass $ \mu=(r_0 M)^{-1/2}  \sim M^{-2/3} f_0^{1/3}$. It can be detailed calculated as
\begin{align}\label{eq:boson_mass}\nonumber
    \mu =& ~3.46 \times 10^{-13} \, \mathrm{eV} \\
    &\times\left(\frac{1}{1+q}\right)^{\frac{1}{6}}  \left(\frac{30 \, M_\odot}{M}\right)^{\frac{2}{3}} \left(\frac{(1+z) f_0}{1 \, \mathrm{Hz}}\right)^{\frac{1}{3}} ~ .
\end{align}
As a benchmark example to illustrate the observable $D$–$f$ relation in GWs, we consider a superradiant boson cloud surrounding a $30\,M_\odot$ BH. A binary companion with mass ratio $q=1$ is orbiting around it. We compute the evolution of the GW observables $h$, $f$, and $\dd f/\dd t$ from $0.1\,\text{Hz}$ to $10\,\text{Hz}$ for this system, within which the effect of DF dominates over GW emission. We obtain the corresponding $D$–$f$ diagram using Eq.~\eqref{eq:observable_D}, as shown in Fig.~\ref{fig:D_BC}.

The solid lines in Fig.~\ref{fig:D_BC} are the numerical results of $ D-f $ relation in the presence of superradiant cloud. The cloud generates a bumpy feature in the $D-f$ diagram. 
As the interested regime is high GW frequency regime, we use the small $ kr $ approximation $ C_{\Lambda} \propto f^{-2/3} $ in Eq.~\eqref{eq:D_boson_cloud} to demonstrate an analytical-approximated $D-f$ relation as the dashed curve. Since $kr = \alpha \sqrt{ (1+q)r/M } $, a smaller value of $\alpha$ can produce a better $C_\Lambda$ approximation.
By fitting it to the numerical results, we can find a best-fit value of $f_0$. 
Then, we can obtain the boson mass with Eq.~\eqref{eq:boson_mass} in which the BH mass, mass ratio, and redshift would be extracted from the GW waveform in the $ P_{\rm GW}$-dominated frequency range. 
Note that the ground-based and the future space-based GW detectors can cover both the $ P_{\rm DF} $-dominated and the $ P_{\rm GW} $-dominated frequency bands such that our method is applicable to realistic GW analysis.

We also estimate the $|v_r/v_\phi|$ in superradiant cloud to check the validity of adiabatic orbital evolution approximation. For $\alpha = 0.1$, $|v_r/v_\phi| < 0.23$ holds for entire frequency range in Fig.~\ref{fig:D_BC} and $|v_r/v_\phi| < 0.05$ for $f > 0.5 \, \text{Hz}$, hence this adiabatic orbital evolution approximation is satisfied.

To end this part, we place some comments on our treatment. 
We have used the small $ kr $ approximation to obtain a power-law form of $ C_{\Lambda} $ when performing the fitting of $ D $. 
In principle, the fitting at the corresponding high-frequency regime should be more accurate than the low-frequency part. 
However, when we fit and obtain $ \mu $ in Fig.~\ref{fig:D_BC}, we have adopted the global best-fit value among $0.1$ Hz to $10$ Hz for simplicity.
The resulting $ \mu $ is about 5\% different from the true value we used for the solid lines, which is acceptable for our current purpose.

\subsection{Probing DM Halo via GWs}

\begin{figure}[htbp]
	\centering
	\includegraphics[width=8cm]{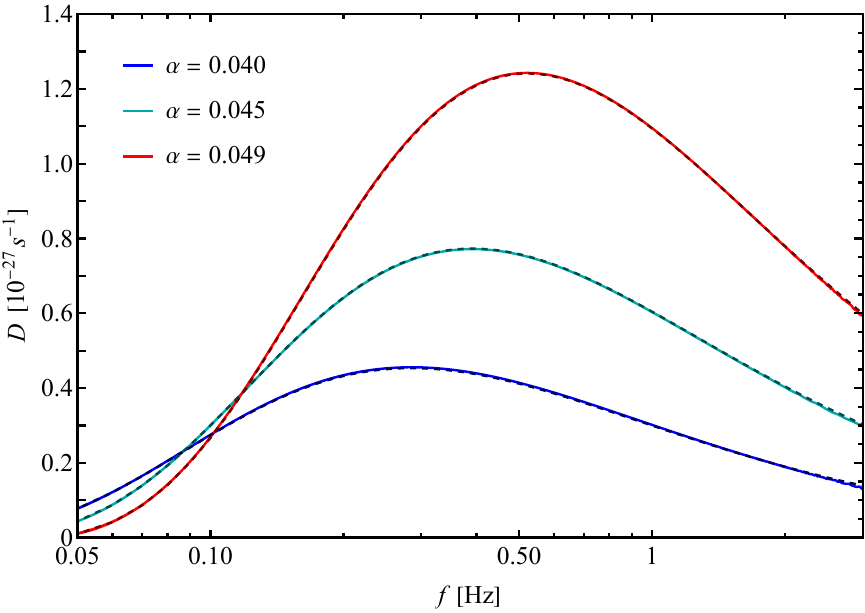}
	\caption{The $D-f$ diagram for a soliton with $\alpha = 0.04, 0.045,0.049$ that corresponds to boson mass $\mu = 1.8, 2,2.2 \times10^{-13} \, \text{eV}$, respectively. The solid curves represent the numerical results obtained using $C_\Lambda$ calculated from Eq.~\eqref{Eq.ULBCLambda}. The dashed curves are the best-fit analytical approximation of $D$ by assuming $C_\Lambda \sim f^{-2/3}$. We set the mass ratio of solition and the BH is $\varepsilon = 10^{-3}$, parameters of the BH binary as $M = 30 \, M_\odot$ and $q = 1$, its corresponding redshift is set as $z = 0.5$.}
	\label{fig:D_Soliton}
\end{figure}

For ultralight bosons surrounding a Schwarzschild BH, a soliton-like structure forms with a density profile given by Eq.~\eqref{eq:soliton_density_profile}. 
The corresponding quantity $D$ for the soliton can then be expressed as 
\begin{align}
    D  &= \frac{8G}{c^4d_L(z)} |P_{\rm DF}|\left(\frac{11}{2} - 2\left(\frac{f_0}{f}\right)^{\frac{2}{3}} - \frac{3}{2} \frac{\dd \ln C_\Lambda}{\dd \ln f} \right) \\\nonumber
    &\propto \left(\frac{f_0}{f}\right)^{\frac{1}{3}} e^{-2\left(\frac{f_0}{f}\right)^{\frac{2}{3}}}\left(\frac{11}{2} - 2\left(\frac{f_0}{f}\right)^{\frac{2}{3}} - \frac{3}{2} \frac{\dd \ln C_\Lambda}{\dd \ln f} \right) C_\Lambda\,.
\end{align}
We consider a soliton formed around a $30 \, M_\odot$ BH, the mass ratio between the solitonic core and BH is set as $\varepsilon = 10^{-3}$. 
Then we calculate observable evolution $h$, $f$ and $\dd f/\dd t$ of GWs from the DF dominant frequency region, $0.05\,\text{Hz}$ to $3\,\text{Hz}$, numerically in this binary with a soliton, and use $h$, $f$ and $\dd f/\dd t$ to construct $D$ as defined in Eq.~\eqref{eq:observable_D}. 
We obtain the $D-f$ diagram as shown in Fig.~\ref{fig:D_Soliton}.
The resulting $D$–$f$ diagram exhibits a distinct behavior compared with that of the superradiant boson cloud, because of the different underlying density profiles. 
Similar to what we have discussed in the superradiant case, the boson mass of this soliton-like DM halo can be inferred from the characteristic frequency $f_0$. 
Once $f_0$ is fitted from the observed $D$–$f$ relation, the corresponding boson mass can be obtained using Eq.~\eqref{eq:boson_mass}. 
Here, we have also used the analytical $D-f$ relation obtained with small $ kr $ approximation for the fitting, which results in about 3\% error from the true value of $ \mu $.
Again, the global best-fit value among $0.05$ Hz to $3$ Hz of $ \mu $ is adopted.
We should notice that the perfect match of the dashed curve with numerical result is due to the global fitting in a narrow frequency range. Meanwhile, the smaller value of $\alpha$ in solitons leads to a better fitting result than that in superradiant boson clouds.

In this case, the value of $|v_r/v_\phi|$ is also smaller than $5\times10^{-4}$ within the considered frequency range for $\alpha = 0.04$. 
This ensures the validity of the adiabatic orbital evolution approximation.

\begin{figure}[htbp]
	\centering
	\includegraphics[width=8cm]{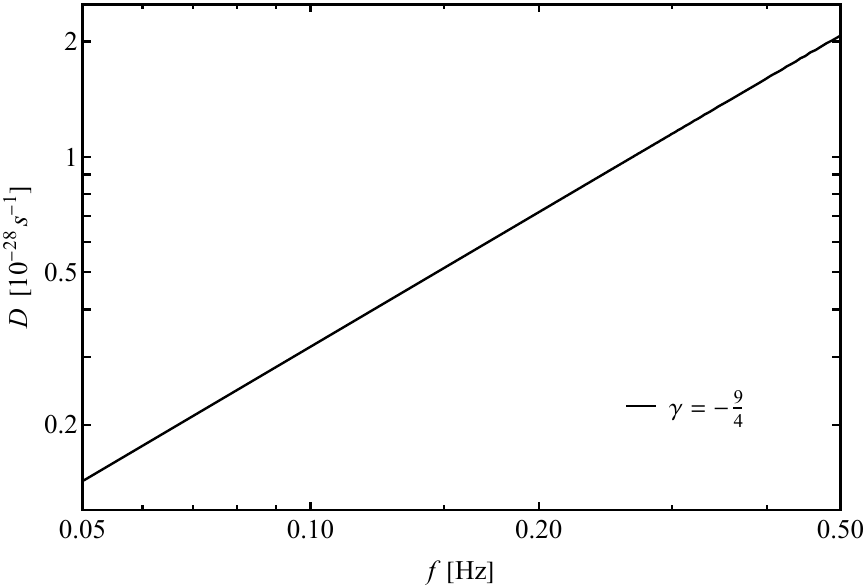}
	\caption{The $D-f$ diagram for a dark matter halo with power index $\gamma = -9/4$. We set the parameters of the BH binary as $M = 30 \, M_\odot$ and $q = 1$, the corresponding redshift is set as $z = 0.5$.}
	\label{fig:D_DMH}
\end{figure}

For particle-like CDM, on the other hand, the density profile of formed halos follows $\rho \propto r^\gamma$, and $C_\Lambda \sim \text{const}$, so Eq.~\eqref{eq:observable_D} can be rewritten as
\begin{align}\label{eq:D_DM_halo}\nonumber
    D &= \frac{8G}{c^4d_L(z)} |P_{\rm DF}|\left(\gamma+ \frac{11}{2} \right) \\
    &\propto f^{-\frac{2}{3}(\gamma+\frac{1}{2})} \left(\gamma + \frac{11}{2}\right)~.
\end{align}
As a benchmark example, we consider such a halo surrounding a BH of $M=30\,M_\odot$ with a binary companion to illustrate the $D$–$f$ relation. We compute the evolution of the GW observables $h$, $f$, and $\dd f/\dd t$ from the DF dominate frequency region, $0.05\,\text{Hz}$ to $0.5\,\text{Hz}$, for this binary system, and construct $D$ using Eq.~\eqref{eq:observable_D}. The resulting $D$–$f$ diagram is shown in Fig.~\ref{fig:D_DMH}. 

Eq.~\eqref{eq:D_DM_halo} shows that the power-law index of the $D$–$f$ relation is directly correlated with the power-law index $\gamma$ in the density profile of the dark environment, as
\begin{align}\label{eq:DMH_power_index}
    \frac{\dd \ln D}{\dd \ln f} = -\frac{2}{3}(\gamma+\frac{1}{2})~.
\end{align}
Once such a $D$–$f$ relation is observed in GWs, it can reveal the CDM halo distribution surrounding the BH binary, and may indicate a possible primordial origin of the central BH.
In a spike-like halo with a power-index $\gamma = -9/4$, $|v_r/v_\phi| < 6 \times 10^{-5}$ holds in DF-dominated frequency, which ensures the validity of the adiabatic orbital evolution approximation.

It should be noted that measurement uncertainties in the GW waveform can significantly affect the inferred $D$–$f$ diagram. If the GW emission power dominates over the DF power loss in the orbital frequency evolution, the waveform deviations induced by DF will be smaller than the observational uncertainties. In such cases, the $D$–$f$ relation cannot be reliably extracted from the GW signal. Therefore, to obtain a measurable $D$–$f$ diagram, one should focus on the frequency range where the DF power loss dominates over the GW emission power. A detailed discussion of this point is provided in the next section.

\section{Detectability of dark dense environment in GWs}\label{sec:DM_detectability}

To probe the features of dark dense environments via GW waveforms, the signals must first be detectable by GW detectors. 
For a GW event to be confidently detected, its signal-to-noise ratio (SNR) must exceed a given threshold within the observation time. Following Refs.~\cite{Rosado:2015voo, Ding:2020ykt}, we define the optimal SNR as
\begin{align}
    \text{SNR} = \sqrt{4 \int_{f_{\rm min}}^{f_{\rm max}} \frac{|\tilde{h}(f)|^2}{S_n(f)} \dd f}~,
\end{align}
with a Fourier transform of GW waveform $\tilde{h}(f)$ as
\begin{align}
    \tilde{h}(f) = \sqrt{\frac{5}{24}} \frac{(G \mathcal{M}_c (1+z))^{5/6}}{\pi^{2/3} c^{3/2} d_L(z)} f^{-7/6}~.
\end{align}
Here, $f_{\rm min}$ and $f_{\rm max}$ denote the initial and final observed GW frequencies during the observation period which is taken to be one year in the following analysis. $S_n(f)$ represents the detector’s noise strain
\cite{Moore:2014lga}. To ensure a detection probability greater than $95\%$, corresponding to a false-alarm probability below $0.1\%$, the optimal SNR must exceed a conservative threshold value given by \cite{LIGOScientific:2016vbw}
\begin{align}
    \text{SNR} > 8~.
\end{align}
It should be noted that the detection method described above is based on matching of GW templates~\cite{Ajith:2007kx} which are modeled for binary systems without considering dark dense environments. 
Therefore, detecting GWs influenced by such environments requires template-independent approaches, such as sine-Gaussian wavelets \cite{Cornish:2014kda, LIGOScientific:2016fbo} or deep-learning \cite{George:2017pmj}. 
These template-independent methods have demonstrated performance comparable to the template-based approach in the case of GW150914 \cite{LIGOScientific:2016aoc} and are expected to be improved further. 
Consequently, we adopt the same detection criterion of $\text{SNR} > 8$ used in template-based analyses for GWs both with and without dark dense environments.

After detecting the GW waveform, in order to use the new quantity $D$ for probing the dense dark environment around BHs, we must focus on the GW frequency range where $D$ is measurable. 
As discussed at the end of Sec.~\ref{sec:GW_probe}, for $D$ to be observable, the dissipative power governing the evolution of the GW waveform must be dominated by DF, ensuring that the corresponding signal is not buried in the GW noise. 
Therefore, we consider
\begin{align}\label{eq:DF_GW_power}
    P_{\rm DF} > 10 \, P_{\rm GW} ~ ,
\end{align}
as a criterion that the DF effect would not be hidden in the GW noise.
We can then illustrate the range of DM parameters that can be probed through the quantity $D$ in GW waveforms. 
Conversely, if no such features in $D$ is detected within the corresponding frequency range, the associated DM parameter space can be excluded.

\subsection{Detectability of GA Boson Cloud}

\begin{figure}[htbp]
	\centering
	\includegraphics[width=8cm]{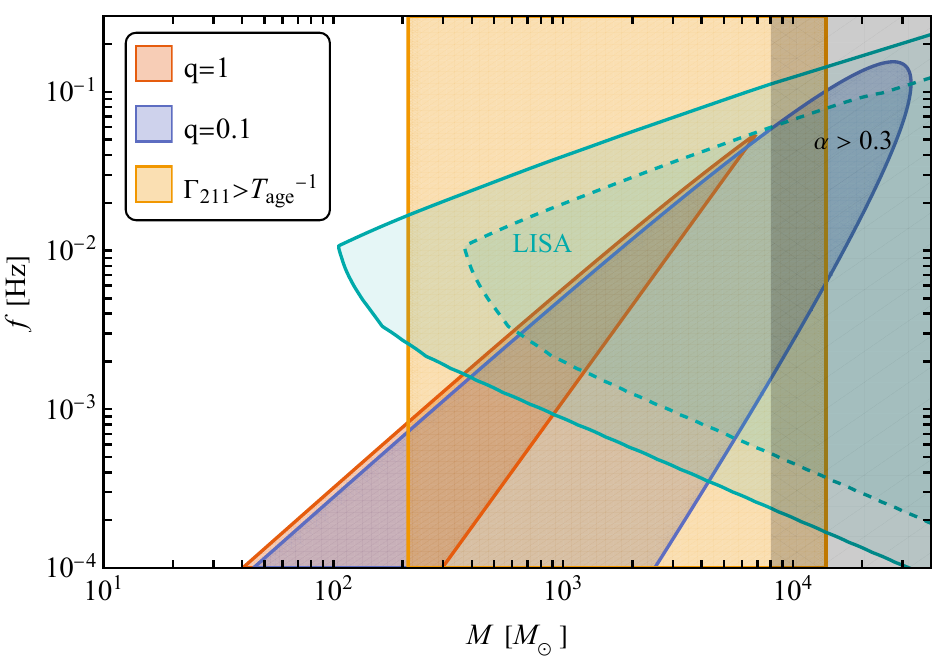}
	\caption{The illustration of the parameter space where the power loss due to DF dominates over that from GW by a factor of 10 for $q=1,\,0.1$. We take $\mu = 5 \times 10^{-15}$ eV as our benchmark bosonic mass and fix the redshift at $z = 0.5$. The region with $\xi > 1$ is excluded to ensure the validity of our analysis. The yellow region indicates where the superradiant growth rate of the $211$ cloud exceeds the inverse age of the universe, implying that the cloud can form within the age of the universe. The shaded region on the right denotes $\alpha > 0.3$. The cyan shadow regions correspond to the detectable parameter regions in LISA \cite{bender1998lisa} with an observation time of one year. The solid cyan curve corresponds to the case $q=1$, while the dashed one corresponds to the case $q=0.1$. }
	\label{fig:211MBf}
\end{figure}

We first consider the scenario of a GA boson cloud surrounding a BH in a binary system. By applying Eq.~\eqref{eq:DF_GW_power} to the superradiant boson cloud model with an ultralight boson mass of $\mu = 5 \times 10^{-15}\,\text{eV}$, we obtain the detectable frequency ranges for different BH masses $M$, as shown in Fig.~\ref{fig:211MBf}.

In this figure, we do not show the case with $q = 0.01$ since it is nearly identical to the $q = 0.1$ case. 
We also see that the feasible region for $q = 1$ is much smaller than that for $q = 0.1$. 
Actually, the mass ratio has only a weak influence on the ratio between $P_{\rm GW}$ and $P_{\rm DF}$. 
The reduced region in the $q = 1$ case arises primarily from the exclusion imposed by the condition $\xi > 1$, under which Eq.~\eqref{Eq.ULBCLambda} is no longer valid. 
The yellow parameter region on the right is determined by the constraint on the superradiant rate, $\Gamma_{211} > T_{\rm age}^{-1}$, where $\Gamma_{211}$ is defined in Eq.~\eqref{Eq.SuperRadRate} and $T_{\rm age}$ denotes the age of the Universe. 
This condition ensures that the GA can form through superradiant instability within the cosmic lifetime. 
However, it is worth noting that a GA may also form via accretion or relaxation of DM \cite{Budker:2023sex}, in which case this constraint can be relaxed. 
The shaded region on the right denotes $\alpha > 0.3$, where the analytical expression for the superradiant rate is no longer valid, as discussed in Sect.~\ref{Sec.SuperradiantBackground}. 
The solid cyan curve corresponds to the case $q=1$, while the dashed one corresponds to the case $q=0.1$. 
As an example of the conclusion that can be drawn from this figure, the feasible frequency range in which $P_{\rm DF}$ can be detected through the quantity $D$ is $(0.001,\,0.06)\,\text{Hz}$ for $q=1$ and a boson mass of $\mu = 5\times10^{-15}\,\text{eV}$, assuming that the cloud is purely generated by superradiance. 
This frequency region lies inside the detectable region in LISA with an observation time of one year. 

\begin{figure}[htbp]
	\centering
	\includegraphics[width=8cm]{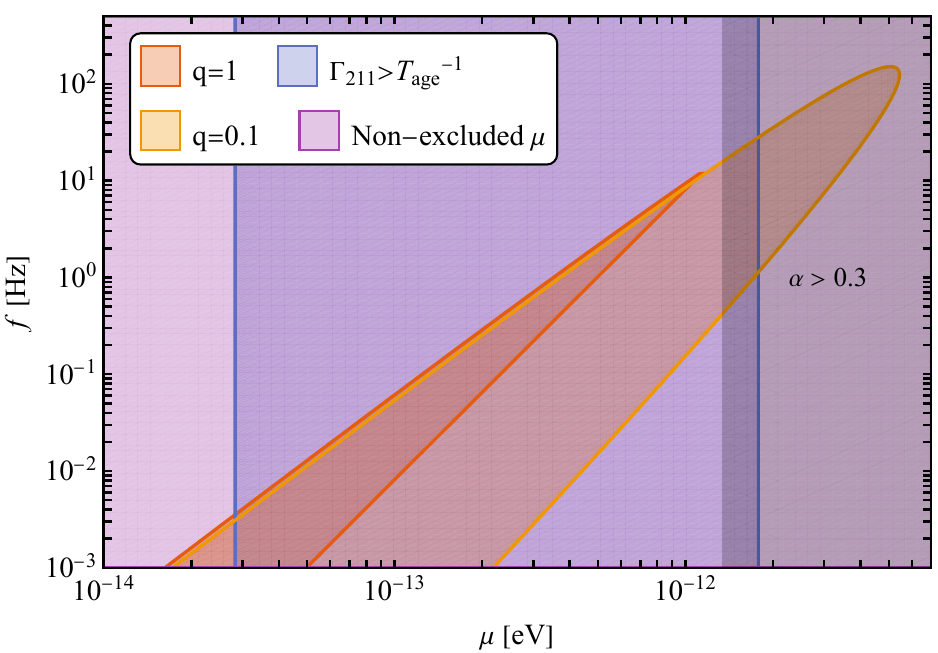}
    \includegraphics[width=8.2cm]{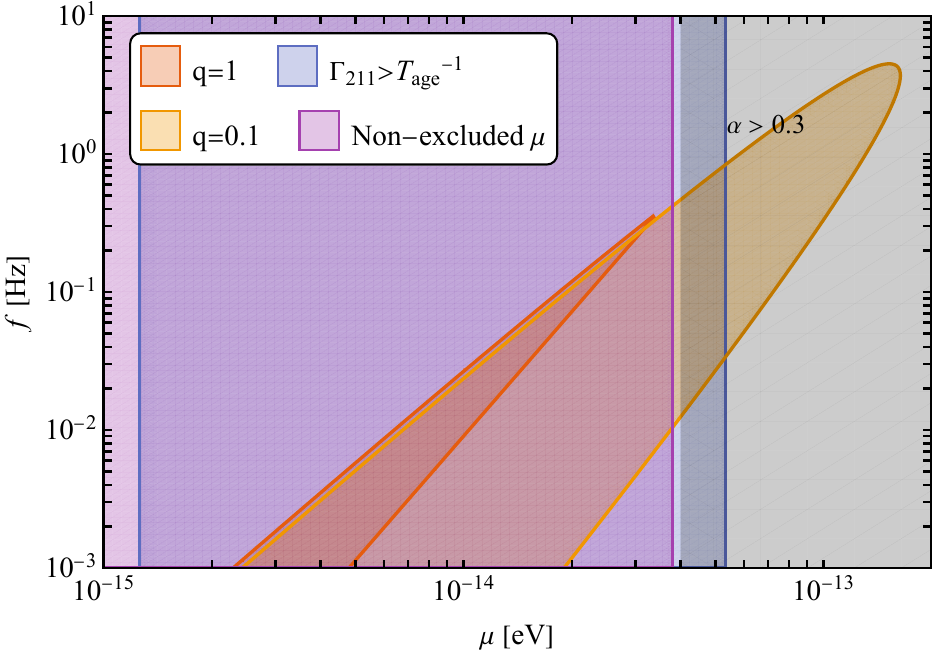}
	\caption{The illustration of the parameter space where the power loss due to DF dominates over that from GW with $q=1,\,0.1$ by a factor of 10. We take $M_B=30M_\odot$ in the top panel and $M_B=1000 M_\odot$ for the bottom panel as our benchmark BH mass, and fix the redshift at $z = 0.5$. The region with $\xi > 1$ is excluded to ensure the validity of our analysis. The blue region indicates where the superradiant growth rate of the $211$ cloud exceeds the inverse age of the universe, implying that the cloud can form within the age of the universe. The purple shaded region corresponds to the scalar bosonic mass range that has not been excluded in \cite{Stott:2020gjj}. The shaded region on the right denotes $\alpha > 0.3$.}
	\label{fig:211muf}
\end{figure}

We also present the detectable parameter regions for different boson masses $\mu$ and GW frequencies $f$, with a fixed BH mass, in Fig.~\ref{fig:211muf}. 
In the top panel, the primary BH mass is set to $M = 30\,M_{\odot}$, while in the bottom panel it is $M = 1000\,M_{\odot}$. 
The shaded region on the right corresponds to $\alpha > 0.3$, where the analytical expression for the superradiant rate becomes invalid, as discussed in Sec.~\ref{Sec.SuperradiantBackground}. 
The purple shaded regions on the left of each plot indicate the non-excluded boson mass ranges reported in Ref.~\cite{Stott:2020gjj}. 
The exclusion arises from BH spin observations, since the superradiant instability extracts angular momentum from the BH, thereby preventing the existence of highly spinning BHs in the Regge plane. 
Consequently, the observation of high-spin BHs implies that the superradiant instability has not occurred, meaning that bosons within the corresponding mass range do not exist. 
As a result, for $M= 30\,M_{\odot}$, most of the parameter space is ruled out by this constraint, whereas for $M = 1000\,M_{\odot}$, the situation is significantly relaxed, leaving a much larger viable parameter region.

Another point to be clarified is that, although we present the results for $q = 1$ in all cases, several subtleties arise in the equal-mass binary system. For instance, the DF may become strong enough to disrupt or scatter the cloud. Also, the cloud can hardly survive the influence of the binary companion, as it may undergo either resonant transitions or off-resonant couplings to states that are rapidly absorbed by the BH. These backreaction of the binary on cloud would decrease the amplitude of $D$ and modify the $D-f$ relation. In addition, a large $M_\ast$ results in a larger $\xi$, which is more likely to exceed unity. This would violate the condition $\xi \ll 1$, thereby breaking the validity of Eq.~\eqref{Eq.ULBCLambda} that we have used throughout this work and changing the shape of $D-f$ diagram.

\subsection{Detectability of DM Halo}

\begin{figure}[htbp]
	\centering
	\includegraphics[width=7.7cm]{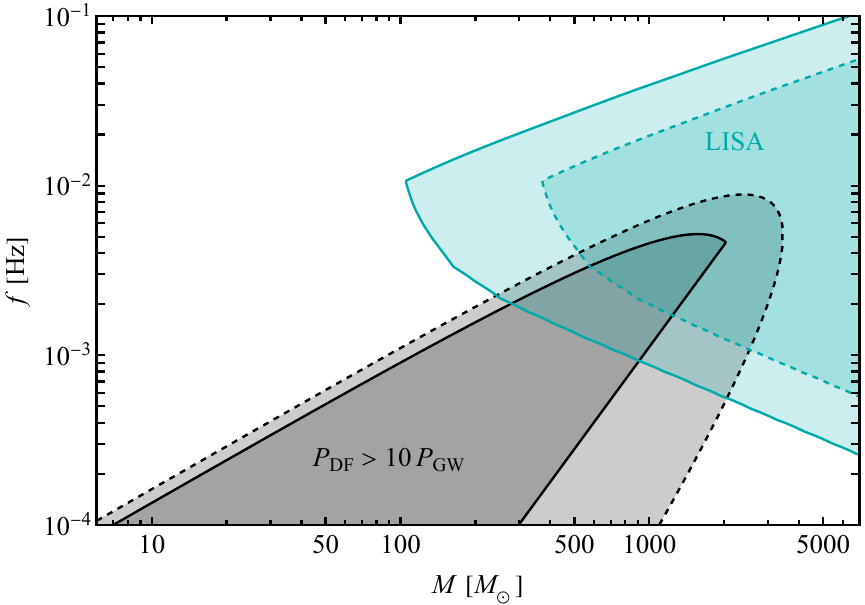}
        \includegraphics[width=8cm]{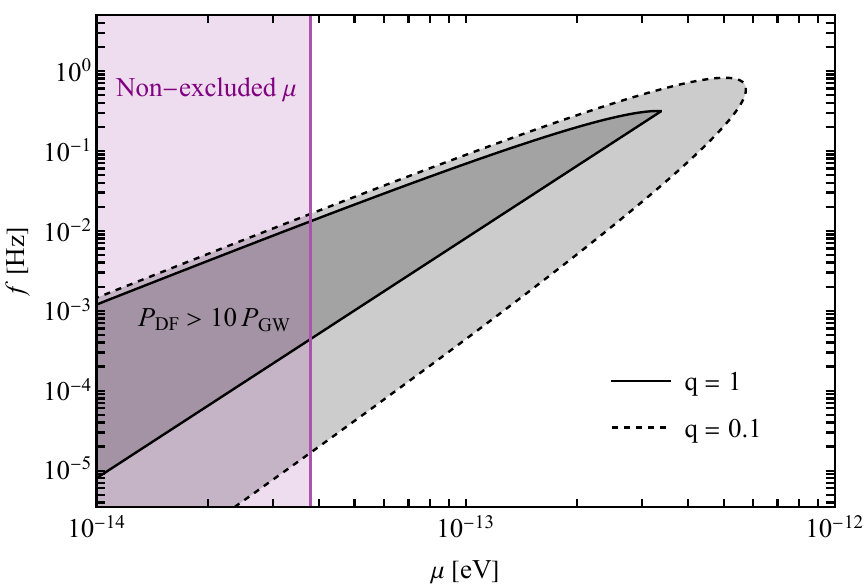}
	\caption{[Top] The illustration of the detection space for soliton, where the power loss due to DF dominates over that from GW by a factor of 10. We take the boson mass $\mu = 5\times10^{-15} \, \text{eV}$. The cyan shadow regions correspond with detectable parameter regions in LISA \cite{bender1998lisa} with an observation time of one year. [Bottom] The illustration of the detection space for the soliton, where the power loss due to DF dominates over that from GW by a factor of 10. The BH mass is set as $M = 30 \, M_\odot$. On both top and bottom panels, we fix redshift at $z = 0.5$ and set the BH mass ratio $q = 1, 0.1$ for solid and dashed curves, respectively. The region with $\xi > 1$ is excluded to ensure the validity of our analysis. The purple shaded region corresponds to the scalar bosonic mass range that has not been excluded in \cite{Stott:2020gjj}.}
	\label{fig:Soliton_Mf}
\end{figure}

For the soliton surrounding a Schwarzschild BH, we compute the DF power using Eq.~\eqref{eq:soliton_density_profile} and show the region satisfying $P_{\rm DF} > 10\,P_{\rm GW}$ in Fig.~\ref{fig:Soliton_Mf}. 
In the top panel of Fig.~\ref{fig:Soliton_Mf}, we consider a fixed boson mass of $\mu = 5\times10^{-15}\,\text{eV}$ around BHs of various masses and evaluate their detectability with the space-based GW detector LISA, adopting the criterion $\text{SNR} > 8$. 
 As an example for the conclusion we can draw from this figure,
for a binary with primary BH of $1000\,M_\odot$ and $q = 1$, the detectable frequency range of a soliton-like DM halo lies between $(0.001,\, 0.005)\,\text{Hz}$. 
In the bottom panel of Fig.~\ref{fig:Soliton_Mf}, we consider different boson masses with fixed BH mass $M = 30\,M_\odot$ and show the corresponding frequency range where the DF power dominates.

\begin{figure}[htbp]
	\centering
	\includegraphics[width=8cm]{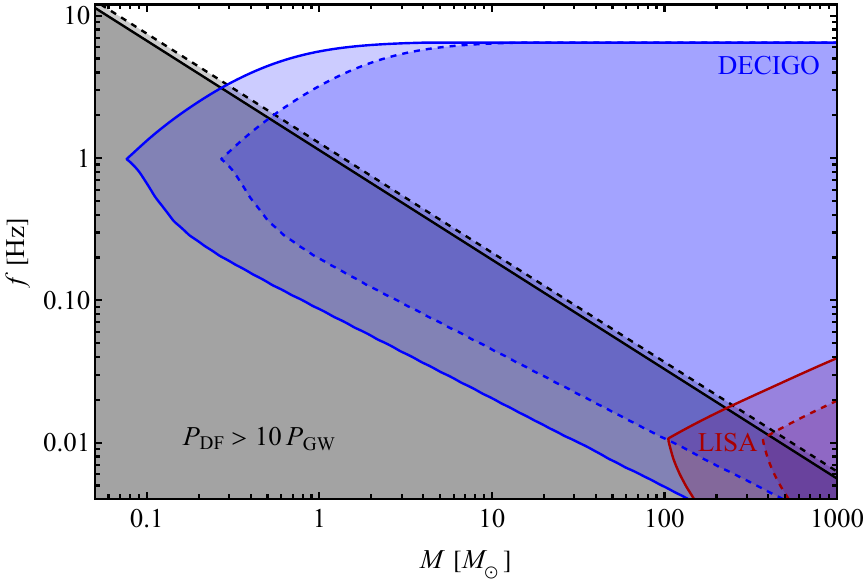}
	\caption{The illustration of the detection space where the power loss due to DF dominates over that from GW by a factor of 10. We take $\gamma=-9/4$ as our benchmark DM halo profile, and fix redshift at $z = 0.5$. We set the BH mass ratio $q = 1, 0.1$ for solid and dashed curve, respectively. The blue and red shadow regions correspond with detectable parameter regions in DECIGO \cite{Kawamura:2006up, Kawamura:2020pcg} and LISA \cite{bender1998lisa}, respectively. We set an observation time as one year.}
	\label{fig:DMH_Mf}
\end{figure}

For a halo formed by particle-like CDM surrounding a binary BH, the density profile follows $\rho \propto r^{\gamma}$, leading to a DF power scaling as $P_{\rm DF} \propto f^{-(2\gamma + 1)/3}$, while the GW emission power scales as $P_{\rm GW} \propto f^{10/3}$. 
This implies that the DF power dominates the orbital evolution of binary BHs at low frequencies when the CDM halo has a power index of $\gamma = -9/4$. 
Using Eqs.~\eqref{eq:DF_power} and \eqref{eq:GW_power}, we show the frequency range satisfying $P_{\rm DF} > 10\,P_{\rm GW}$ in Fig.~\ref{fig:DMH_Mf}.

The results show that the such a halo dominates the GW waveform evolution in the low-frequency regime as expected. 
Such imprints of the halo on GW waveforms should be detectable by future GW detectors. 
In Fig.~\ref{fig:DMH_Mf}, we illustrate the detectable frequency ranges for the space-based detectors DECIGO and LISA, corresponding to $\text{SNR} > 8$ for a one-year observation. 
For a $30\,M_\odot$ BH binary, the influence of such type of halos becomes significant at $f < 0.08\,\text{Hz}$, and its imprint can be detected by DECIGO in the frequency range $(0.011,\, 0.083)\,\text{Hz}$, where a characteristic power-law $D$–$f$ relation is expected to appear. 
For more massive BH binaries, the halo–dominated region shifts to lower frequencies, making the signal accessible to LISA and even lower-frequency GW detectors such as SKA. We also show the case with a mass ratio $q = 0.1$ as dashed curves in Fig.~\ref{fig:DMH_Mf}, which exhibits only a slight shift in the detectable frequency range compared to the $q = 1$ case.

\section{Discussions and Conclusions}\label{sec:conclusion}

To conclude, we have investigated the imprints of dark dense environments on GW waveforms. Using the GW observables such as the amplitude $h$, frequency $f$, and its time derivative $\dd f/\dd t$, we have constructed a novel quantity, $D$, which is proportional to the DF power loss contributed by the surrounding dark environment. The evolution of $D$ in the GW frequency domain yields a $D$–$f$ diagram, which characterizes the properties of density profiles of the dark dense environments surrounding BH binaries.

We analyzed the quantity $D$ in three types of dark dense environments: a GA boson cloud, a soliton-like condensate, and a compact mini halo consists of particle-like CDM. 
For the GA boson cloud, fitting the theoretical prediction of $D$ to the observed $D$–$f$ diagram helps us to determine the characteristic GW frequency $ f_0 $ which gives the Bohr radius $ r_0 $ of the GA.
Combined with the BH mass, mass ratio, and redshift, which are quantities measurable in the GW power–dominated regime, we can extract the boson mass via Eq.~\eqref{eq:boson_mass}. 
Ultralight bosons can form soliton-like halos around Schwarzschild BHs, and the boson mass can similarly be inferred by fitting the $D$–$f$ relation. 
On the other hand, particle-like CDM forms spike-like halos whose density profile index $\gamma$ can be determined from the power-law index of the observed $D$–$f$ diagram, as given in Eq.~\eqref{eq:DMH_power_index}. 
It is worth noticing that, astrophysical stellar-mass BHs are unlikely to produce the third type environment.
Therefore, the detection of a power-law $D$–$f$ behavior would strongly suggest a primordial origin of these BHs. 
Lastly, we have investigated the detectability of dark dense environments with space-based GW detectors such as DECIGO and LISA. 
Our results show that a wide DF-dominated frequency range lies within their sensitivity bands, allowing the extraction of DM properties through $D$. 
Conversely, a null detection in this regime would place stringent constraints on the corresponding DM parameters space.

Our work raises a number of interesting follow-up questions that we can explore in the future. For example, we focus exclusively on BH binaries with circular orbits in this work, since a nonzero eccentricity would introduce an additional $\dd e/\dd t$ term into the analytical expression of $D$, complicating the analysis. 
Although various mechanisms during the inspiral phase tend to circularize the orbit \cite{Hinder:2007qu, Franciolini:2021xbq}, the assumption of $e = 0$ reduces the number of BH binary events that can be used to study dark dense environments through the $D$–$f$ diagram. 
Furthermore, detecting GW waveforms modified by DF requires the establishment of template-independent analysis techniques \cite{Cornish:2014kda, LIGOScientific:2016fbo, George:2017pmj, Akhshi:2020xmd, Shimomura:2025fhz}, as current template-based methods \cite{Ajith:2007kx} are inadequate for our purpose. 
Therefore, the development of template-independent GW analysis is essential to apply our method to effectively probe new physics around BH binaries.
We leave these topics to future works.

The dark dense environments surrounding BHs make them ideal laboratories for exploring the nature of DM. 
With the advancement of space-based GW detectors, the GW frontier will extend into lower-frequency and higher-sensitivity regimes \cite{Gair:2022knq}. 
This will enable precise measurements of the $D$–$f$ diagram in GW waveforms during the inspiral phase, offering an insight of the dark dense environments around BHs and deepening our understanding of DM.

\section*{Acknowledgments}
The authors would like to thank Jun'ichi Yokoyama and Josu Aurrekoetxea for the useful discussion and Misao Sasaki for helpful comments. Q.D. and H.Y.Z. were supported by IBS under the project code, IBS-R018-D3. M.H. was supported by IBS under the project code, IBS-R018-D1.

\appendix

\section{Consistency Check for Adiabatic Orbital Evolution Approximation}\label{Sec.selfconsistency}

In this appendix, we derive the self-consistency criterion 
for the adiabatic orbital shrinking approximation.

First, we decompose the relative velocity into its radial and 
angular components $ \vec{v} = v_r \hat{r} + v_\phi \hat{\phi} $ in the rest frame of the primary BH. 
The radial component is calculated as 
\begin{align}
    v_r &= \frac{\dd r}{\dd t_s} = -\frac{2}{3} \frac{r}{f_s} \frac{\dd f_s}{\dd t_s} \\
    &= -\frac{2}{3} \frac{\left( GM \right)^{1/3} (1+q)^{1/3}}{\pi^{2/3}} \frac{1}{f_s^{5/3}} \frac{\dd f_s}{\dd t_s} ~,
\end{align}
where we use Kepler's law for circular orbit while deriving the third equality. 
On the other hand, the angular component reads 
\begin{align}
    v_\phi \simeq \sqrt{\frac{GM (1+q)}{r}} = \left( \pi GM (1+q) f_s\right)^{1/3} ~. 
\end{align}
The adiabatic approximation is valid as long as the radial velocity component is much 
smaller than the tangential one, that is 
\begin{align}\label{eq:vel_ratio}
    \left| \frac{v_r}{v_\phi} \right| = \frac{2}{3\pi}\frac{1}{f^2} \frac{\dd f}{\dd t} \ll 1 ~. 
\end{align}
As the prefactor is order unity, we simply require 
\begin{align}
    \frac{1}{f^2} \frac{\dd f}{\dd t} \ll 1 ~.
\end{align}
Specifically, in our consideration with $ P_{\rm GW} $ and $ P_{\rm DF} $, we can derive 
\begin{align}
    \label{Eq.Consistency}
    \frac{1}{f^2} \frac{\dd f}{\dd t} = \frac{12\sqrt{2}\pi}{5} q\sqrt{1+q} \left( \frac{r_s}{r} \right)^{\frac{5}{2}} \left( 1+ \left| \frac{P_{\rm DF}}{P_{\rm GW}} \right| \right)~,
\end{align}
where $r_s$ is the Schwarzschild radius defined as $r_s \equiv 2 G M/c^2$.
Qualitatively, we mainly focus on the regime where the DF is strong enough to be observable, i.e., $|P_{\rm DF}/P_{\rm GW}| \gg 1$. 
Thus, for $ q\sim 1 $ case, 
the adiabatic orbital evolution approximation remains valid only when the DF is significant while the binary separation is still much larger than the Schwarzschild radius of the primary BH. However, for small mass ratio $q$, the constraint on $r$ for which the adiabatic approximation holds becomes less restrictive.

We also use Eq.~\eqref{eq:vel_ratio} to numerically evaluate $|v_r / v_\phi|$ and confirm that, within the frequency ranges considered in this work, the condition $|v_r / v_\phi| \ll 1$ always holds for all three types of dark dense environments. This validates the use of the adiabatic approximation throughout the parameter space analyzed. The corresponding numerical details are presented in the main text.

\bibliographystyle{utphys}
\bibliography{GW_Probe}

\providecommand{\href}[2]{#2}\begingroup\raggedright\begin{thebibliography}{10}

\bibitem{LIGOScientific:2016aoc}
{\bfseries LIGO Scientific, Virgo} Collaboration, B.~P. Abbott {\em et~al.},
  ``{Observation of Gravitational Waves from a Binary Black Hole Merger},''
  \href{http://dx.doi.org/10.1103/PhysRevLett.116.061102}{{\em Phys. Rev.
  Lett.} {\bfseries 116} no.~6, (2016) 061102},
  \href{http://arxiv.org/abs/1602.03837}{{\ttfamily arXiv:1602.03837 [gr-qc]}}.

\bibitem{LIGOScientific:2020tif}
{\bfseries LIGO Scientific, Virgo} Collaboration, R.~Abbott {\em et~al.},
  ``{Tests of general relativity with binary black holes from the second
  LIGO-Virgo gravitational-wave transient catalog},''
  \href{http://dx.doi.org/10.1103/PhysRevD.103.122002}{{\em Phys. Rev. D}
  {\bfseries 103} no.~12, (2021) 122002},
  \href{http://arxiv.org/abs/2010.14529}{{\ttfamily arXiv:2010.14529 [gr-qc]}}.

\bibitem{Isi:2019aib}
M.~Isi, M.~Giesler, W.~M. Farr, M.~A. Scheel, and S.~A. Teukolsky, ``{Testing
  the no-hair theorem with GW150914},''
  \href{http://dx.doi.org/10.1103/PhysRevLett.123.111102}{{\em Phys. Rev.
  Lett.} {\bfseries 123} no.~11, (2019) 111102},
  \href{http://arxiv.org/abs/1905.00869}{{\ttfamily arXiv:1905.00869 [gr-qc]}}.

\bibitem{Isi:2020tac}
M.~Isi, W.~M. Farr, M.~Giesler, M.~A. Scheel, and S.~A. Teukolsky, ``{Testing
  the Black-Hole Area Law with GW150914},''
  \href{http://dx.doi.org/10.1103/PhysRevLett.127.011103}{{\em Phys. Rev.
  Lett.} {\bfseries 127} no.~1, (2021) 011103},
  \href{http://arxiv.org/abs/2012.04486}{{\ttfamily arXiv:2012.04486 [gr-qc]}}.

\bibitem{Tang:2025jyj}
S.-P. Tang, H.-T. Wang, Y.-J. Li, and Y.-Z. Fan, ``{Verification of the Black
  Hole Area Law with GW230814},''
  \href{http://arxiv.org/abs/2509.03480}{{\ttfamily arXiv:2509.03480 [gr-qc]}}.

\bibitem{KAGRA:2025oiz}
{\bfseries KAGRA, Virgo, LIGO Scientific} Collaboration, A.~G. Abac {\em
  et~al.}, ``{GW250114: Testing Hawking{\textquoteright}s Area Law and the Kerr
  Nature of Black Holes},'' \href{http://dx.doi.org/10.1103/kw5g-d732}{{\em
  Phys. Rev. Lett.} {\bfseries 135} no.~11, (2025) 111403},
  \href{http://arxiv.org/abs/2509.08054}{{\ttfamily arXiv:2509.08054 [gr-qc]}}.

\bibitem{Schutz:1986gp}
B.~F. Schutz, ``{Determining the Hubble Constant from Gravitational Wave
  Observations},'' \href{http://dx.doi.org/10.1038/323310a0}{{\em Nature}
  {\bfseries 323} (1986) 310--311}.

\bibitem{Holz:2005df}
D.~E. Holz and S.~A. Hughes, ``{Using gravitational-wave standard sirens},''
  \href{http://dx.doi.org/10.1086/431341}{{\em Astrophys. J.} {\bfseries 629}
  (2005) 15--22}, \href{http://arxiv.org/abs/astro-ph/0504616}{{\ttfamily
  arXiv:astro-ph/0504616}}.

\bibitem{LIGOScientific:2017vwq}
{\bfseries LIGO Scientific, Virgo} Collaboration, B.~P. Abbott {\em et~al.},
  ``{GW170817: Observation of Gravitational Waves from a Binary Neutron Star
  Inspiral},'' \href{http://dx.doi.org/10.1103/PhysRevLett.119.161101}{{\em
  Phys. Rev. Lett.} {\bfseries 119} no.~16, (2017) 161101},
  \href{http://arxiv.org/abs/1710.05832}{{\ttfamily arXiv:1710.05832 [gr-qc]}}.

\bibitem{Ding:2023smy}
Q.~Ding, ``{Merger rate of primordial black hole binaries as a probe of the
  Hubble parameter},''
  \href{http://dx.doi.org/10.1103/PhysRevD.110.063542}{{\em Phys. Rev. D}
  {\bfseries 110} no.~6, (2024) 063542},
  \href{http://arxiv.org/abs/2312.13728}{{\ttfamily arXiv:2312.13728
  [astro-ph.CO]}}.

\bibitem{Chen:2024gdn}
H.-Y. Chen, J.~M. Ezquiaga, and I.~Gupta, ``{Cosmography with next-generation
  gravitational wave detectors},''
  \href{http://dx.doi.org/10.1088/1361-6382/ad424f}{{\em Class. Quant. Grav.}
  {\bfseries 41} no.~12, (2024) 125004},
  \href{http://arxiv.org/abs/2402.03120}{{\ttfamily arXiv:2402.03120 [gr-qc]}}.

\bibitem{Clesse:2020ghq}
S.~Clesse and J.~Garcia-Bellido, ``{GW190425, GW190521 and GW190814: Three
  candidate mergers of primordial black holes from the QCD epoch},''
  \href{http://dx.doi.org/10.1016/j.dark.2022.101111}{{\em Phys. Dark Univ.}
  {\bfseries 38} (2022) 101111},
  \href{http://arxiv.org/abs/2007.06481}{{\ttfamily arXiv:2007.06481
  [astro-ph.CO]}}.

\bibitem{DeLuca:2020sae}
V.~De~Luca, V.~Desjacques, G.~Franciolini, P.~Pani, and A.~Riotto, ``{GW190521
  Mass Gap Event and the Primordial Black Hole Scenario},''
  \href{http://dx.doi.org/10.1103/PhysRevLett.126.051101}{{\em Phys. Rev.
  Lett.} {\bfseries 126} no.~5, (2021) 051101},
  \href{http://arxiv.org/abs/2009.01728}{{\ttfamily arXiv:2009.01728
  [astro-ph.CO]}}.

\bibitem{Yuan:2025avq}
C.~Yuan, Z.-C. Chen, and L.~Liu, ``{GW231123 Mass Gap Event and the Primordial
  Black Hole Scenario},'' \href{http://arxiv.org/abs/2507.15701}{{\ttfamily
  arXiv:2507.15701 [astro-ph.CO]}}.

\bibitem{Kokkotas:1999bd}
K.~D. Kokkotas and B.~G. Schmidt, ``{Quasinormal modes of stars and black
  holes},'' \href{http://dx.doi.org/10.12942/lrr-1999-2}{{\em Living Rev. Rel.}
  {\bfseries 2} (1999) 2}, \href{http://arxiv.org/abs/gr-qc/9909058}{{\ttfamily
  arXiv:gr-qc/9909058}}.

\bibitem{Berti:2009kk}
E.~Berti, V.~Cardoso, and A.~O. Starinets, ``{Quasinormal modes of black holes
  and black branes},''
  \href{http://dx.doi.org/10.1088/0264-9381/26/16/163001}{{\em Class. Quant.
  Grav.} {\bfseries 26} (2009) 163001},
  \href{http://arxiv.org/abs/0905.2975}{{\ttfamily arXiv:0905.2975 [gr-qc]}}.

\bibitem{Konoplya:2011qq}
R.~A. Konoplya and A.~Zhidenko, ``{Quasinormal modes of black holes: From
  astrophysics to string theory},''
  \href{http://dx.doi.org/10.1103/RevModPhys.83.793}{{\em Rev. Mod. Phys.}
  {\bfseries 83} (2011) 793--836},
  \href{http://arxiv.org/abs/1102.4014}{{\ttfamily arXiv:1102.4014 [gr-qc]}}.

\bibitem{Chen:2020lpq}
X.~Chen, Z.-Y. Xuan, and P.~Peng, ``{Fake massive black holes in the
  milli-Hertz gravitational-wave band},''
  \href{http://dx.doi.org/10.3847/1538-4357/ab919f}{{\em Astrophys. J.}
  {\bfseries 896} no.~2, (2020) 171},
  \href{http://arxiv.org/abs/2003.08639}{{\ttfamily arXiv:2003.08639
  [astro-ph.HE]}}.

\bibitem{Takatsy:2025bfk}
J.~Tak{\'a}tsy, L.~Zwick, K.~Hendriks, P.~Saini, G.~Fabj, and J.~Samsing,
  ``{The construction and use of dephasing prescriptions for environmental
  effects in gravitational wave astronomy},''
  \href{http://arxiv.org/abs/2505.09513}{{\ttfamily arXiv:2505.09513
  [astro-ph.HE]}}.

\bibitem{Gondolo:1999ef}
P.~Gondolo and J.~Silk, ``{Dark matter annihilation at the galactic center},''
  \href{http://dx.doi.org/10.1103/PhysRevLett.83.1719}{{\em Phys. Rev. Lett.}
  {\bfseries 83} (1999) 1719--1722},
  \href{http://arxiv.org/abs/astro-ph/9906391}{{\ttfamily
  arXiv:astro-ph/9906391}}.

\bibitem{Eda:2013gg}
K.~Eda, Y.~Itoh, S.~Kuroyanagi, and J.~Silk, ``{New Probe of Dark-Matter
  Properties: Gravitational Waves from an Intermediate-Mass Black Hole Embedded
  in a Dark-Matter Minispike},''
  \href{http://dx.doi.org/10.1103/PhysRevLett.110.221101}{{\em Phys. Rev.
  Lett.} {\bfseries 110} no.~22, (2013) 221101},
  \href{http://arxiv.org/abs/1301.5971}{{\ttfamily arXiv:1301.5971 [gr-qc]}}.

\bibitem{Bar:2019pnz}
N.~Bar, K.~Blum, T.~Lacroix, and P.~Panci, ``{Looking for ultralight dark
  matter near supermassive black holes},''
  \href{http://dx.doi.org/10.1088/1475-7516/2019/07/045}{{\em JCAP} {\bfseries
  07} (2019) 045}, \href{http://arxiv.org/abs/1905.11745}{{\ttfamily
  arXiv:1905.11745 [astro-ph.CO]}}.

\bibitem{Mack:2006gz}
K.~J. Mack, J.~P. Ostriker, and M.~Ricotti, ``{Growth of structure seeded by
  primordial black holes},'' \href{http://dx.doi.org/10.1086/518998}{{\em
  Astrophys. J.} {\bfseries 665} (2007) 1277--1287},
  \href{http://arxiv.org/abs/astro-ph/0608642}{{\ttfamily
  arXiv:astro-ph/0608642}}.

\bibitem{Akil:2023kym}
A.~Akil and Q.~Ding, ``{A dark matter probe in accreting pulsar-black hole
  binaries},'' \href{http://dx.doi.org/10.1088/1475-7516/2023/09/011}{{\em
  JCAP} {\bfseries 09} (2023) 011},
  \href{http://arxiv.org/abs/2304.08824}{{\ttfamily arXiv:2304.08824
  [astro-ph.HE]}}.

\bibitem{Kadota:2023wlm}
K.~Kadota, J.~H. Kim, P.~Ko, and X.-Y. Yang, ``{Gravitational wave probes on
  self-interacting dark matter surrounding an intermediate mass black hole},''
  \href{http://dx.doi.org/10.1103/PhysRevD.109.015022}{{\em Phys. Rev. D}
  {\bfseries 109} no.~1, (2024) 015022},
  \href{http://arxiv.org/abs/2306.10828}{{\ttfamily arXiv:2306.10828
  [hep-ph]}}.

\bibitem{Aurrekoetxea:2023jwk}
J.~C. Aurrekoetxea, K.~Clough, J.~Bamber, and P.~G. Ferreira, ``{Effect of Wave
  Dark Matter on Equal Mass Black Hole Mergers},''
  \href{http://dx.doi.org/10.1103/PhysRevLett.132.211401}{{\em Phys. Rev.
  Lett.} {\bfseries 132} no.~21, (2024) 211401},
  \href{http://arxiv.org/abs/2311.18156}{{\ttfamily arXiv:2311.18156 [gr-qc]}}.

\bibitem{Ding:2024mro}
Q.~Ding, M.~He, and V.~Takhistov, ``{Primordial Black Hole Mergers as Probes of
  Dark Matter in the Galactic Center},''
  \href{http://dx.doi.org/10.3847/1538-4357/adaeb4}{{\em Astrophys. J.}
  {\bfseries 981} no.~1, (2025) 62},
  \href{http://arxiv.org/abs/2410.02591}{{\ttfamily arXiv:2410.02591
  [astro-ph.CO]}}.

\bibitem{Tomaselli:2025zdo}
G.~M. Tomaselli and A.~Caputo, ``{Probing dense environments around Sgr A* with
  S-stars dynamics},'' \href{http://arxiv.org/abs/2509.03568}{{\ttfamily
  arXiv:2509.03568 [astro-ph.GA]}}.

\bibitem{Spieksma:2025exm}
T.~F.~M. Spieksma, ``{Exploring Black Hole Environments},'' other thesis, 9,
  2025.

\bibitem{Roy:2025qaa}
S.~Roy, R.~Vicente, J.~C. Aurrekoetxea, K.~Clough, and P.~G. Ferreira,
  ``{Scalar fields around black hole binaries in LIGO-Virgo-KAGRA},''
  \href{http://arxiv.org/abs/2510.17967}{{\ttfamily arXiv:2510.17967 [gr-qc]}}.

\bibitem{Eda:2014kra}
K.~Eda, Y.~Itoh, S.~Kuroyanagi, and J.~Silk, ``{Gravitational waves as a probe
  of dark matter minispikes},''
  \href{http://dx.doi.org/10.1103/PhysRevD.91.044045}{{\em Phys. Rev. D}
  {\bfseries 91} no.~4, (2015) 044045},
  \href{http://arxiv.org/abs/1408.3534}{{\ttfamily arXiv:1408.3534 [gr-qc]}}.

\bibitem{Ghodla:2024fit}
S.~Ghodla, ``{Galactic and extragalactic probe of dark matter with LISA's
  binary black holes},''
  \href{http://dx.doi.org/10.1088/1475-7516/2025/02/036}{{\em JCAP} {\bfseries
  02} (2025) 036}, \href{http://arxiv.org/abs/2410.15562}{{\ttfamily
  arXiv:2410.15562 [astro-ph.CO]}}.

\bibitem{Arvanitaki:2009fg}
A.~Arvanitaki, S.~Dimopoulos, S.~Dubovsky, N.~Kaloper, and J.~March-Russell,
  ``{String Axiverse},''
  \href{http://dx.doi.org/10.1103/PhysRevD.81.123530}{{\em Phys. Rev. D}
  {\bfseries 81} (2010) 123530},
  \href{http://arxiv.org/abs/0905.4720}{{\ttfamily arXiv:0905.4720 [hep-th]}}.

\bibitem{Arvanitaki:2010sy}
A.~Arvanitaki and S.~Dubovsky, ``{Exploring the String Axiverse with Precision
  Black Hole Physics},''
  \href{http://dx.doi.org/10.1103/PhysRevD.83.044026}{{\em Phys. Rev. D}
  {\bfseries 83} (2011) 044026},
  \href{http://arxiv.org/abs/1004.3558}{{\ttfamily arXiv:1004.3558 [hep-th]}}.

\bibitem{Brito:2015oca}
R.~Brito, V.~Cardoso, and P.~Pani, ``{Superradiance}: {New Frontiers in Black
  Hole Physics},'' \href{http://dx.doi.org/10.1007/978-3-319-19000-6}{{\em
  Lect. Notes Phys.} {\bfseries 906} (2015) pp.1--237},
  \href{http://arxiv.org/abs/1501.06570}{{\ttfamily arXiv:1501.06570 [gr-qc]}}.

\bibitem{Baumann:2019eav}
D.~Baumann, H.~S. Chia, J.~Stout, and L.~ter Haar, ``{The Spectra of
  Gravitational Atoms},''
  \href{http://dx.doi.org/10.1088/1475-7516/2019/12/006}{{\em JCAP} {\bfseries
  12} (2019) 006}, \href{http://arxiv.org/abs/1908.10370}{{\ttfamily
  arXiv:1908.10370 [gr-qc]}}.

\bibitem{Baumann:2019ztm}
D.~Baumann, H.~S. Chia, R.~A. Porto, and J.~Stout, ``{Gravitational Collider
  Physics},'' \href{http://dx.doi.org/10.1103/PhysRevD.101.083019}{{\em Phys.
  Rev. D} {\bfseries 101} no.~8, (2020) 083019},
  \href{http://arxiv.org/abs/1912.04932}{{\ttfamily arXiv:1912.04932 [gr-qc]}}.

\bibitem{Ding:2020bnl}
Q.~Ding, X.~Tong, and Y.~Wang, ``{Gravitational Collider Physics via
  Pulsar-Black Hole Binaries},''
  \href{http://dx.doi.org/10.3847/1538-4357/abd803}{{\em Astrophys. J.}
  {\bfseries 908} no.~1, (2021) 78},
  \href{http://arxiv.org/abs/2009.11106}{{\ttfamily arXiv:2009.11106
  [astro-ph.HE]}}.

\bibitem{Takahashi:2021yhy}
T.~Takahashi, H.~Omiya, and T.~Tanaka, ``{Axion cloud evaporation during
  inspiral of black hole binaries: The effects of backreaction and
  radiation},'' \href{http://dx.doi.org/10.1093/ptep/ptac044}{{\em PTEP}
  {\bfseries 2022} no.~4, (2022) 043E01},
  \href{http://arxiv.org/abs/2112.05774}{{\ttfamily arXiv:2112.05774 [gr-qc]}}.

\bibitem{Tong:2021whq}
X.~Tong, Y.~Wang, and H.-Y. Zhu, ``{Gravitational Collider Physics via
  Pulsar\textendash{}Black Hole Binaries II: Fine and Hyperfine Structures Are
  Favored},'' \href{http://dx.doi.org/10.3847/1538-4357/ac36db}{{\em Astrophys.
  J.} {\bfseries 924} no.~2, (2022) 99},
  \href{http://arxiv.org/abs/2106.13484}{{\ttfamily arXiv:2106.13484
  [astro-ph.HE]}}.

\bibitem{Tong:2022bbl}
X.~Tong, Y.~Wang, and H.-Y. Zhu, ``{Termination of superradiance from a binary
  companion},'' \href{http://dx.doi.org/10.1103/PhysRevD.106.043002}{{\em Phys.
  Rev. D} {\bfseries 106} no.~4, (2022) 043002},
  \href{http://arxiv.org/abs/2205.10527}{{\ttfamily arXiv:2205.10527 [gr-qc]}}.

\bibitem{Baumann:2021fkf}
D.~Baumann, G.~Bertone, J.~Stout, and G.~M. Tomaselli, ``{Ionization of
  gravitational atoms},''
  \href{http://dx.doi.org/10.1103/PhysRevD.105.115036}{{\em Phys. Rev. D}
  {\bfseries 105} no.~11, (2022) 115036},
  \href{http://arxiv.org/abs/2112.14777}{{\ttfamily arXiv:2112.14777 [gr-qc]}}.

\bibitem{Takahashi:2023flk}
T.~Takahashi, H.~Omiya, and T.~Tanaka, ``{Evolution of binary systems
  accompanying axion clouds in extreme mass ratio inspirals},''
  \href{http://dx.doi.org/10.1103/PhysRevD.107.103020}{{\em Phys. Rev. D}
  {\bfseries 107} no.~10, (2023) 103020},
  \href{http://arxiv.org/abs/2301.13213}{{\ttfamily arXiv:2301.13213 [gr-qc]}}.

\bibitem{Tomaselli:2023ysb}
G.~M. Tomaselli, T.~F.~M. Spieksma, and G.~Bertone, ``{Dynamical friction in
  gravitational atoms},''
  \href{http://dx.doi.org/10.1088/1475-7516/2023/07/070}{{\em JCAP} {\bfseries
  07} (2023) 070}, \href{http://arxiv.org/abs/2305.15460}{{\ttfamily
  arXiv:2305.15460 [gr-qc]}}.

\bibitem{Fan:2023jjj}
K.~Fan, X.~Tong, Y.~Wang, and H.-Y. Zhu, ``{Modulating binary dynamics via the
  termination of black hole superradiance},''
  \href{http://dx.doi.org/10.1103/PhysRevD.109.024059}{{\em Phys. Rev. D}
  {\bfseries 109} no.~2, (2024) 024059},
  \href{http://arxiv.org/abs/2311.17013}{{\ttfamily arXiv:2311.17013 [gr-qc]}}.

\bibitem{Zhu:2024bqs}
H.-Y. Zhu, X.~Tong, G.~Manzoni, and Y.~Ma, ``{Survival of the Fittest: Testing
  Superradiance Termination with Simulated Binary Black Hole Statistics},''
  \href{http://dx.doi.org/10.3847/1538-4357/adb1db}{{\em Astrophys. J.}
  {\bfseries 981} no.~2, (2025) 165},
  \href{http://arxiv.org/abs/2409.14159}{{\ttfamily arXiv:2409.14159 [gr-qc]}}.

\bibitem{Dosopoulou:2023umg}
F.~Dosopoulou, ``{Dynamical friction in dark matter spikes: Corrections to
  Chandrasekhar{\textquoteright}s formula},''
  \href{http://dx.doi.org/10.1103/PhysRevD.110.083027}{{\em Phys. Rev. D}
  {\bfseries 110} no.~8, (2024) 083027},
  \href{http://arxiv.org/abs/2305.17281}{{\ttfamily arXiv:2305.17281
  [astro-ph.HE]}}.

\bibitem{Ding:2025nxe}
Q.~Ding, M.~He, V.~Takhistov, and H.-Y. Zhu, ``{Dark Matter-Independent Orbital
  Decay Bounds on Ultralight Bosons from OJ287},''  (5, 2025) ,
  \href{http://arxiv.org/abs/2505.09696}{{\ttfamily arXiv:2505.09696
  [hep-ph]}}.

\bibitem{Gnedin:2003rj}
O.~Y. Gnedin and J.~R. Primack, ``{Dark Matter Profile in the Galactic
  Center},'' \href{http://dx.doi.org/10.1103/PhysRevLett.93.061302}{{\em Phys.
  Rev. Lett.} {\bfseries 93} (2004) 061302},
  \href{http://arxiv.org/abs/astro-ph/0308385}{{\ttfamily
  arXiv:astro-ph/0308385}}.

\bibitem{Ricotti:2007au}
M.~Ricotti, J.~P. Ostriker, and K.~J. Mack, ``{Effect of Primordial Black Holes
  on the Cosmic Microwave Background and Cosmological Parameter Estimates},''
  \href{http://dx.doi.org/10.1086/587831}{{\em Astrophys. J.} {\bfseries 680}
  (2008) 829}, \href{http://arxiv.org/abs/0709.0524}{{\ttfamily arXiv:0709.0524
  [astro-ph]}}.

\bibitem{Berezinsky:2013fxa}
V.~S. Berezinsky, V.~I. Dokuchaev, and Y.~N. Eroshenko, ``{Formation and
  internal structure of superdense dark matter clumps and ultracompact
  minihaloes},'' \href{http://dx.doi.org/10.1088/1475-7516/2013/11/059}{{\em
  JCAP} {\bfseries 11} (2013) 059},
  \href{http://arxiv.org/abs/1308.6742}{{\ttfamily arXiv:1308.6742
  [astro-ph.CO]}}.

\bibitem{Oguri:2022fir}
M.~Oguri, V.~Takhistov, and K.~Kohri, ``{Revealing dark matter dress of
  primordial black holes by cosmological lensing},''
  \href{http://dx.doi.org/10.1016/j.physletb.2023.138276}{{\em Phys. Lett. B}
  {\bfseries 847} (2023) 138276},
  \href{http://arxiv.org/abs/2208.05957}{{\ttfamily arXiv:2208.05957
  [astro-ph.CO]}}.

\bibitem{GilChoi:2023ahp}
H.~Gil~Choi, S.~Jung, P.~Lu, and V.~Takhistov, ``{Coexistence Test of
  Primordial Black Holes and Particle Dark Matter from Diffractive Lensing},''
  \href{http://dx.doi.org/10.1103/PhysRevLett.133.101002}{{\em Phys. Rev.
  Lett.} {\bfseries 133} no.~10, (2024) 101002},
  \href{http://arxiv.org/abs/2311.17829}{{\ttfamily arXiv:2311.17829
  [astro-ph.CO]}}.

\bibitem{Brax:2019npi}
P.~Brax, J.~A.~R. Cembranos, and P.~Valageas, ``{Fate of scalar dark matter
  solitons around supermassive galactic black holes},''
  \href{http://dx.doi.org/10.1103/PhysRevD.101.023521}{{\em Phys. Rev. D}
  {\bfseries 101} no.~2, (2020) 023521},
  \href{http://arxiv.org/abs/1909.02614}{{\ttfamily arXiv:1909.02614
  [astro-ph.CO]}}.

\bibitem{Hui:2019aqm}
L.~Hui, D.~Kabat, X.~Li, L.~Santoni, and S.~S.~C. Wong, ``{Black Hole Hair from
  Scalar Dark Matter},''
  \href{http://dx.doi.org/10.1088/1475-7516/2019/06/038}{{\em JCAP} {\bfseries
  06} (2019) 038}, \href{http://arxiv.org/abs/1904.12803}{{\ttfamily
  arXiv:1904.12803 [gr-qc]}}.

\bibitem{Planck:2018vyg}
{\bfseries Planck} Collaboration, N.~Aghanim {\em et~al.}, ``{Planck 2018
  results. VI. Cosmological parameters},''
  \href{http://dx.doi.org/10.1051/0004-6361/201833910}{{\em Astron. Astrophys.}
  {\bfseries 641} (2020) A6}, \href{http://arxiv.org/abs/1807.06209}{{\ttfamily
  arXiv:1807.06209 [astro-ph.CO]}}. [Erratum: Astron.Astrophys. 652, C4
  (2021)].

\bibitem{Tsukada:2018mbp}
L.~Tsukada, T.~Callister, A.~Matas, and P.~Meyers, ``{First search for a
  stochastic gravitational-wave background from ultralight bosons},''
  \href{http://dx.doi.org/10.1103/PhysRevD.99.103015}{{\em Phys. Rev. D}
  {\bfseries 99} no.~10, (2019) 103015},
  \href{http://arxiv.org/abs/1812.09622}{{\ttfamily arXiv:1812.09622
  [astro-ph.HE]}}.

\bibitem{1943ApJ....97..255C}
S.~{Chandrasekhar}, ``{Dynamical Friction. I. General Considerations: the
  Coefficient of Dynamical Friction.},''
  \href{http://dx.doi.org/10.1086/144517}{{\em Astrophys. J.} {\bfseries 97}
  (Mar., 1943) 255}.

\bibitem{Hui:2016ltb}
L.~Hui, J.~P. Ostriker, S.~Tremaine, and E.~Witten, ``{Ultralight scalars as
  cosmological dark matter},''
  \href{http://dx.doi.org/10.1103/PhysRevD.95.043541}{{\em Phys. Rev. D}
  {\bfseries 95} no.~4, (2017) 043541},
  \href{http://arxiv.org/abs/1610.08297}{{\ttfamily arXiv:1610.08297
  [astro-ph.CO]}}.

\bibitem{Teukolsky:1973ha}
S.~A. Teukolsky, ``{Perturbations of a rotating black hole. 1. Fundamental
  equations for gravitational electromagnetic and neutrino field
  perturbations},'' \href{http://dx.doi.org/10.1086/152444}{{\em Astrophys. J.}
  {\bfseries 185} (1973) 635--647}.

\bibitem{Leaver:1985ax}
E.~W. Leaver, ``{An Analytic representation for the quasi normal modes of Kerr
  black holes},'' \href{http://dx.doi.org/10.1098/rspa.1985.0119}{{\em Proc.
  Roy. Soc. Lond. A} {\bfseries 402} (1985) 285--298}.

\bibitem{Detweiler:1980uk}
S.~L. Detweiler, ``{KLEIN-GORDON EQUATION AND ROTATING BLACK HOLES},''
  \href{http://dx.doi.org/10.1103/PhysRevD.22.2323}{{\em Phys. Rev. D}
  {\bfseries 22} (1980) 2323--2326}.

\bibitem{Brito2015b}
R.~Brito, V.~Cardoso, and P.~Pani, {\em {Superradiance : energy extraction,
  black-hole bombs and implications for astrophysics and particle physics}}.
\newblock Cham: Springer, New York, 2015.

\bibitem{Cannizzaro:2023jle}
E.~Cannizzaro, L.~Sberna, S.~R. Green, and S.~Hollands, ``{Relativistic
  perturbation theory for black-hole boson clouds},''  (9, 2023) ,
  \href{http://arxiv.org/abs/2309.10021}{{\ttfamily arXiv:2309.10021 [gr-qc]}}.

\bibitem{Baumann:2018vus}
D.~Baumann, H.~S. Chia, and R.~A. Porto, ``{Probing Ultralight Bosons with
  Binary Black Holes},''
  \href{http://dx.doi.org/10.1103/PhysRevD.99.044001}{{\em Phys. Rev. D}
  {\bfseries 99} no.~4, (2019) 044001},
  \href{http://arxiv.org/abs/1804.03208}{{\ttfamily arXiv:1804.03208 [gr-qc]}}.

\bibitem{Cao:2023fyv}
Y.~Cao and Y.~Tang, ``{Signatures of ultralight bosons in compact binary
  inspiral and outspiral},''
  \href{http://dx.doi.org/10.1103/PhysRevD.108.123017}{{\em Phys. Rev. D}
  {\bfseries 108} no.~12, (2023) 123017},
  \href{http://arxiv.org/abs/2307.05181}{{\ttfamily arXiv:2307.05181 [gr-qc]}}.

\bibitem{Li:2025qyu}
X.~Li, J.~Ren, and X.-L. Zhang, ``{Probing Boson Clouds with Supermassive Black
  Hole Binaries},'' \href{http://arxiv.org/abs/2505.02866}{{\ttfamily
  arXiv:2505.02866 [hep-ph]}}.

\bibitem{Aghaie:2023lan}
M.~Aghaie, G.~Armando, A.~Dondarini, and P.~Panci, ``{Bounds on ultralight dark
  matter from NANOGrav},''
  \href{http://dx.doi.org/10.1103/PhysRevD.109.103030}{{\em Phys. Rev. D}
  {\bfseries 109} no.~10, (2024) 103030},
  \href{http://arxiv.org/abs/2308.04590}{{\ttfamily arXiv:2308.04590
  [astro-ph.CO]}}.

\bibitem{Navarro:1995iw}
J.~F. Navarro, C.~S. Frenk, and S.~D.~M. White, ``{The Structure of cold dark
  matter halos},'' \href{http://dx.doi.org/10.1086/177173}{{\em Astrophys. J.}
  {\bfseries 462} (1996) 563--575},
  \href{http://arxiv.org/abs/astro-ph/9508025}{{\ttfamily
  arXiv:astro-ph/9508025}}.

\bibitem{Navarro:1996gj}
J.~F. Navarro, C.~S. Frenk, and S.~D.~M. White, ``{A Universal density profile
  from hierarchical clustering},'' \href{http://dx.doi.org/10.1086/304888}{{\em
  Astrophys. J.} {\bfseries 490} (1997) 493--508},
  \href{http://arxiv.org/abs/astro-ph/9611107}{{\ttfamily
  arXiv:astro-ph/9611107}}.

\bibitem{Nishikawa:2017chy}
H.~Nishikawa, E.~D. Kovetz, M.~Kamionkowski, and J.~Silk,
  ``{Primordial-black-hole mergers in dark-matter spikes},''
  \href{http://dx.doi.org/10.1103/PhysRevD.99.043533}{{\em Phys. Rev. D}
  {\bfseries 99} no.~4, (2019) 043533},
  \href{http://arxiv.org/abs/1708.08449}{{\ttfamily arXiv:1708.08449
  [astro-ph.CO]}}.

\bibitem{Boudaud:2021irr}
M.~Boudaud, T.~Lacroix, M.~Stref, J.~Lavalle, and P.~Salati, ``{In-depth
  analysis of the clustering of dark matter particles around primordial black
  holes. Part~I. Density profiles},''
  \href{http://dx.doi.org/10.1088/1475-7516/2021/08/053}{{\em JCAP} {\bfseries
  08} (2021) 053}, \href{http://arxiv.org/abs/2106.07480}{{\ttfamily
  arXiv:2106.07480 [astro-ph.CO]}}.

\bibitem{Peters:1964zz}
P.~C. Peters, ``{Gravitational Radiation and the Motion of Two Point Masses},''
  \href{http://dx.doi.org/10.1103/PhysRev.136.B1224}{{\em Phys. Rev.}
  {\bfseries 136} (1964) B1224--B1232}.

\bibitem{Hinder:2007qu}
I.~Hinder, B.~Vaishnav, F.~Herrmann, D.~Shoemaker, and P.~Laguna,
  ``{Universality and final spin in eccentric binary black hole inspirals},''
  \href{http://dx.doi.org/10.1103/PhysRevD.77.081502}{{\em Phys. Rev. D}
  {\bfseries 77} (2008) 081502},
  \href{http://arxiv.org/abs/0710.5167}{{\ttfamily arXiv:0710.5167 [gr-qc]}}.

\bibitem{Franciolini:2021xbq}
G.~Franciolini, R.~Cotesta, N.~Loutrel, E.~Berti, P.~Pani, and A.~Riotto,
  ``{How to assess the primordial origin of single gravitational-wave events
  with mass, spin, eccentricity, and deformability measurements},''
  \href{http://dx.doi.org/10.1103/PhysRevD.105.063510}{{\em Phys. Rev. D}
  {\bfseries 105} no.~6, (2022) 063510},
  \href{http://arxiv.org/abs/2112.10660}{{\ttfamily arXiv:2112.10660
  [astro-ph.CO]}}.

\bibitem{2007gwte.book.....M}
M.~{Maggiore},
  \href{http://dx.doi.org/10.1093/acprof:oso/9780198570745.001.0001}{{\em
  {Gravitational Waves: Volume 1: Theory and Experiments}}}.
\newblock 2007.

\bibitem{Rosado:2015voo}
P.~A. Rosado, P.~D. Lasky, E.~Thrane, X.~Zhu, I.~Mandel, and A.~Sesana,
  ``{Detectability of Gravitational Waves from High-Redshift Binaries},''
  \href{http://dx.doi.org/10.1103/PhysRevLett.116.101102}{{\em Phys. Rev.
  Lett.} {\bfseries 116} no.~10, (2016) 101102},
  \href{http://arxiv.org/abs/1512.04950}{{\ttfamily arXiv:1512.04950
  [astro-ph.CO]}}.

\bibitem{Ding:2020ykt}
Q.~Ding, ``{Detectability of primordial black hole binaries at high
  redshift},'' \href{http://dx.doi.org/10.1103/PhysRevD.104.043527}{{\em Phys.
  Rev. D} {\bfseries 104} no.~4, (2021) 043527},
  \href{http://arxiv.org/abs/2011.13643}{{\ttfamily arXiv:2011.13643
  [astro-ph.CO]}}.

\bibitem{Moore:2014lga}
C.~J. Moore, R.~H. Cole, and C.~P.~L. Berry, ``{Gravitational-wave sensitivity
  curves},'' \href{http://dx.doi.org/10.1088/0264-9381/32/1/015014}{{\em Class.
  Quant. Grav.} {\bfseries 32} no.~1, (2015) 015014},
  \href{http://arxiv.org/abs/1408.0740}{{\ttfamily arXiv:1408.0740 [gr-qc]}}.

\bibitem{LIGOScientific:2016vbw}
{\bfseries LIGO Scientific, Virgo} Collaboration, B.~P. Abbott {\em et~al.},
  ``{GW150914: First results from the search for binary black hole coalescence
  with Advanced LIGO},''
  \href{http://dx.doi.org/10.1103/PhysRevD.93.122003}{{\em Phys. Rev. D}
  {\bfseries 93} no.~12, (2016) 122003},
  \href{http://arxiv.org/abs/1602.03839}{{\ttfamily arXiv:1602.03839 [gr-qc]}}.

\bibitem{Ajith:2007kx}
P.~Ajith {\em et~al.}, ``{A Template bank for gravitational waveforms from
  coalescing binary black holes. I. Non-spinning binaries},''
  \href{http://dx.doi.org/10.1103/PhysRevD.77.104017}{{\em Phys. Rev. D}
  {\bfseries 77} (2008) 104017},
  \href{http://arxiv.org/abs/0710.2335}{{\ttfamily arXiv:0710.2335 [gr-qc]}}.
  [Erratum: Phys.Rev.D 79, 129901 (2009)].

\bibitem{Cornish:2014kda}
N.~J. Cornish and T.~B. Littenberg, ``{BayesWave: Bayesian Inference for
  Gravitational Wave Bursts and Instrument Glitches},''
  \href{http://dx.doi.org/10.1088/0264-9381/32/13/135012}{{\em Class. Quant.
  Grav.} {\bfseries 32} no.~13, (2015) 135012},
  \href{http://arxiv.org/abs/1410.3835}{{\ttfamily arXiv:1410.3835 [gr-qc]}}.

\bibitem{LIGOScientific:2016fbo}
{\bfseries LIGO Scientific, Virgo} Collaboration, B.~P. Abbott {\em et~al.},
  ``{Observing gravitational-wave transient GW150914 with minimal
  assumptions},'' \href{http://dx.doi.org/10.1103/PhysRevD.93.122004}{{\em
  Phys. Rev. D} {\bfseries 93} no.~12, (2016) 122004},
  \href{http://arxiv.org/abs/1602.03843}{{\ttfamily arXiv:1602.03843 [gr-qc]}}.
  [Addendum: Phys.Rev.D 94, 069903 (2016)].

\bibitem{George:2017pmj}
D.~George and E.~A. Huerta, ``{Deep Learning for Real-time Gravitational Wave
  Detection and Parameter Estimation: Results with Advanced LIGO Data},''
  \href{http://dx.doi.org/10.1016/j.physletb.2017.12.053}{{\em Phys. Lett. B}
  {\bfseries 778} (2018) 64--70},
  \href{http://arxiv.org/abs/1711.03121}{{\ttfamily arXiv:1711.03121 [gr-qc]}}.

\bibitem{bender1998lisa}
P.~Bender, A.~Brillet, I.~Ciufolini, A.~Cruise, C.~Cutler, K.~Danzmann,
  F.~Fidecaro, W.~Folkner, J.~Hough, P.~McNamara, {\em et~al.}, ``Lisa
  pre-phase a report,'' {\em Max-Planck Institut f{\"u}r Quantenoptik} (1998) .

\bibitem{Budker:2023sex}
D.~Budker, J.~Eby, M.~Gorghetto, M.~Jiang, and G.~Perez, ``{A generic formation
  mechanism of ultralight dark matter solar halos},''
  \href{http://dx.doi.org/10.1088/1475-7516/2023/12/021}{{\em JCAP} {\bfseries
  12} (2023) 021}, \href{http://arxiv.org/abs/2306.12477}{{\ttfamily
  arXiv:2306.12477 [hep-ph]}}.

\bibitem{Stott:2020gjj}
M.~J. Stott, ``{Ultralight Bosonic Field Mass Bounds from Astrophysical Black
  Hole Spin},''  (9, 2020) , \href{http://arxiv.org/abs/2009.07206}{{\ttfamily
  arXiv:2009.07206 [hep-ph]}}.

\bibitem{Kawamura:2006up}
S.~Kawamura {\em et~al.}, ``{The Japanese space gravitational wave antenna
  DECIGO},'' \href{http://dx.doi.org/10.1088/0264-9381/23/8/S17}{{\em Class.
  Quant. Grav.} {\bfseries 23} (2006) S125--S132}.

\bibitem{Kawamura:2020pcg}
S.~Kawamura {\em et~al.}, ``{Current status of space gravitational wave antenna
  DECIGO and B-DECIGO},'' \href{http://dx.doi.org/10.1093/ptep/ptab019}{{\em
  PTEP} {\bfseries 2021} no.~5, (2021) 05A105},
  \href{http://arxiv.org/abs/2006.13545}{{\ttfamily arXiv:2006.13545 [gr-qc]}}.

\bibitem{Akhshi:2020xmd}
A.~Akhshi, H.~Alimohammadi, S.~Baghram, S.~Rahvar, M.~R. Rahimi~Tabar, and
  H.~Arfaei, ``{A template-free approach for waveform extraction of
  gravitational wave events},''
  \href{http://arxiv.org/abs/2005.11352}{{\ttfamily arXiv:2005.11352
  [astro-ph.IM]}}.

\bibitem{Shimomura:2025fhz}
R.~Shimomura, Y.~Tabe, and H.~Shinkai, ``{Gravitational-wave Extraction using
  Independent Component Analysis},''
  \href{http://arxiv.org/abs/2503.14179}{{\ttfamily arXiv:2503.14179 [gr-qc]}}.

\bibitem{Gair:2022knq}
J.~R. Gair, M.~Hewitson, A.~Petiteau, and G.~Mueller, {\em {Space-based
  Gravitational Wave Observatories}}.
\newblock 1, 2022.
\newblock \href{http://arxiv.org/abs/2201.10593}{{\ttfamily arXiv:2201.10593
  [gr-qc]}}.

\end{thebibliography}\endgroup

\end{document}